\documentclass[5p,twocolumn,nonatbib]{elsarticle}

\makeatletter
\def\ps@pprintTitle{\let\@oddhead\@empty\let\@evenhead\@empty
  \let\@oddfoot\@empty\let\@evenfoot\@empty}
\makeatother

\usepackage[T1]{fontenc}
\usepackage[utf8]{inputenc}  
\usepackage{amsmath,amssymb,amsthm,amsfonts} 
\usepackage{graphicx}
\usepackage{float}
\usepackage{stfloats}   
\usepackage{comment}

\usepackage[usenames,dvipsnames,svgnames]{xcolor}
\definecolor{linkpurple}{rgb}{0.3,0,0.9}
\definecolor{blue1}{RGB}{8,24,168}
\definecolor{darkteal}{HTML}{045D5D}
\usepackage[colorlinks=true,linkcolor=blue,citecolor=blue,urlcolor=linkpurple]{hyperref}
\usepackage[
  backend=biber,
  style=phys,
  sorting=none,
  articletitle=false,
  chaptertitle=false,
  biblabel=brackets,
  pageranges=false,
  doi=false,
  url=false,
  eprint=false,
  isbn=false
]{biblatex}
\AtEveryBibitem{%
  \clearfield{title}%
  \ifentrytype{article}{%
    \clearfield{number}%
    \clearfield{issue}%
    \clearfield{month}%
    \clearfield{day}%
  }{}%
}
\AtBeginBibliography{\small}
\DeclareFieldFormat{pages}{#1}
\DeclareFieldFormat{journaltitle}{%
  \iffieldundef{doi}
    {#1}
    {\href{https://doi.org/\thefield{doi}}{#1}}%
}
\renewbibmacro{in:}{} 

\begin{document}

\begin{frontmatter}

 \title{Heavy-quark thermalization in the quark-gluon plasma via a fractional Langevin equation}
\author[a]{Jai Prakash}
\author[b]{Mohammad Yousuf Jamal\corref{cor1}}
\author[c]{Shanshan Cao}
\author[b]{Guang-You Qin\corref{cor1}}
\author[a]{and  Yifeng Sun\corref{cor1}}
\cortext[cor1]{Correspond to: yousufjml5@gmail.com, guangyou.qin@mail.ccnu.edu.cn, sunyfphy@sjtu.edu.cn}
\address[a]{State Key Laboratory of Dark Matter Physics, Shanghai Key Laboratory for Particle Physics and Cosmology, Key Laboratory for Particle Astrophysics and Cosmology (MOE), School of Physics and Astronomy, Shanghai Jiao Tong University, Shanghai 200240, China}
\address[b]{Key Laboratory of Quark and Lepton Physics (MOE) \& Institute of Particle Physics, Central China Normal University, Wuhan 430079, China}
\address[c]{School of Physics, Southeast University, Nanjing 211189, China}
\begin{abstract}
The dynamics of heavy quarks in the quark-gluon plasma is commonly described within a standard Langevin equation framework. In this work, we extend the conventional framework by using a fractional Langevin equation based on the Caputo derivative of order $\alpha$ with $1<\alpha\leq 2$. We solve the fractional Langevin equation and derive the corresponding {\textit {fractional fluctuation-dissipation theorem}} for the heavy quark thermalization in the quark-gluon plasma medium. We calculate the time evolution of the mean-squared momentum, mean-squared displacement, momentum distribution, energy loss, longitudinal momentum loss, transverse momentum broadening, and  the nuclear suppression factor   of the heavy quarks. The mean-squared momentum thermalizes to the same value for all $\alpha$ at late times, in agreement with the fractional fluctuation-dissipation theorem, and the nuclear suppression factor shows a systematic $\alpha$-dependence. The energy loss of an energetic heavy quark grows strongly with $\alpha$. The time evolution of the heavy quark momentum distribution provides additional sensitivity to the fractional dynamics of the heavy quarks in the QGP medium.

\end{abstract}

\end{frontmatter}

\section{Introduction}
\label{sec:intro}
The quark-gluon plasma (QGP) is a state of hot and dense nuclear matter expected to be produced in heavy-ion collisions (HICs)~\cite{STAR:2006vcp,Adams:2005dq,PHENIX:2004vcz,ALICE:2010khr,BRAHMS:2005gow} at the Large Hadron Collider (LHC) and the Relativistic Heavy-Ion Collider (RHIC). The QGP thermalizes and hadronizes on timescales of a few fm/$c$, so direct observation is not possible, and probes that carry information about the medium throughout its space-time evolution are essential for the study of its properties. In this context, heavy quarks (HQs), particularly charm and bottom quarks, have been recognized as effective probes of the QGP medium. They are produced at the very initial stage of the collision, and their larger mass compared to the particles (quarks and gluons) of the medium gives them a thermalization time considerably larger than that of the medium particles~\cite{vanHees:2004gq,Rapp:2009my,Cao:2011et}. They therefore propagate through the medium while retaining memory of their interactions~\cite{Gossiaux:2008jv,PhysRevC.84.044901,Andronic:2015wma,Song:2015sfa,Aarts:2016hap,Cao:2016gvr,Plumari:2017ntm,Dong:2019unq,Plumari:2019hzp,Debnath:2023zet,du2023accelerated,Grishmanovskii:2023gog,Minissale:2023dct,PhysRevLett.130.231902,Sun:2023adv,Bhattacharyya:2024hku,Das:2024vac,Das:2024wht,Oei:2024pva,Oliva:2024rex,ZACCONE2024116483,Das:2025yxy,Grishmanovskii:2025mnc,Sambataro:2025obe,Sambataro:2025pop,Nair:2026tqy}. In the conventional theoretical framework, HQ motion in the QGP is described using the standard Langevin equation~\cite{Greco:2015uhf,vanHees:2007me,PhysRevC.86.034905,Akamatsu:2008ge,PhysRevC.86.014903,Cao:2015hia}, in which the HQ is treated as a Brownian particle in a bath of Gaussian-distributed white noise. Two forces act on the HQ: a dissipative drag and a random force representing the thermal kicks from the medium constituents, related through the fluctuation-dissipation theorem (FDT). The noise correlation is assumed to decay instantaneously, identifying the process as the standard Wiener process. These standard Langevin equations face quantitative limitations in reproducing the measured HQ energy loss in the QGP~\cite{Das:2015ana,Zhang:2022fum}, motivating the search for alternative formulations of the HQ propagation in the medium.

The framework adopted here is the fractional Langevin equation (FLE)~\cite{coffey2012langevin,lutz2001fractional,vitali2018langevin,oliveira2019anomalous,PhysRevE.107.024105,PhysRevE.108.034113,PhysRevLett.125.240606,PhysRevE.104.024115,2017JSP...169..316L}, in which the integer-order time derivative of the standard Langevin equation is replaced by a fractional derivative of order $\alpha$. The FLE generalizes the Brownian dynamics of the HQs through a single parameter $\alpha$ and reduces to the ordinary Langevin equation as $\alpha\to 1$. Its foundational formulation was introduced by Mainardi and collaborators~\cite{MR1611585,mainardi2008fractional,MR1611587}. The resulting transport exhibits anomalous diffusion~\cite{balescu1995anomalous,bouchaud1990anomalous,metzler2000random,PhysRevLett.85.5655,PhysRevLett.104.238102,sokolov2005diffusion}, for which the linear time dependence of the mean-squared displacement characteristic of Einstein's relation no longer holds. In recent work~\cite{Prakash:2024irm,Prakash:2024rdz}, anomalous HQ diffusion was studied within the FLE framework using the standard Einstein fluctuation-dissipation relation. Its validity in the fractional setting is not guaranteed, and is addressed here through the derivation, of a fractional fluctuation-dissipation theorem (F-FDT) directly from the FLE. The F-FDT fixes the relation between the drag and diffusion coefficients for arbitrary $\alpha$ and reduces to the standard FDT in the same $\alpha\to 1$ limit. Among the mathematical tools available for constructing fractional equations of motion, the Caputo fractional derivative is the most widely employed in the Langevin formalism~\cite{guo2013numerics,li2013finite,miller1993introduction}; alternative formulations~\cite{carpinteri2014fractals,Agrawal_2007,Feller1971} are not used here. Within this framework, the time evolution of the mean-squared momentum, $\langle p^{2}(t)\rangle$, the mean-squared displacement, $\langle x^{2}(t)\rangle$, the energy loss, $\langle\Delta E\rangle$, the longitudinal momentum loss, $\langle\Delta p_{\parallel}\rangle$, the transverse momentum broadening, $\langle p_{\perp}^{2}\rangle$, the nuclear suppression factor, $R_{AA}$, the HQ momentum distribution, $dN/dp_{T}$, and a path-length estimate of the elliptic flow, $v_{2}$, are computed, yielding a self-consistent assessment of the fractional dynamics on HQ transport in the QGP.

The paper is organized as follows. Section~\ref{sec:FLE} introduces the FLE for heavy quarks. Section~\ref{sec:FFDT} derives the F-FDT. Section~\ref{sec:results} presents the numerical results. The final section~\ref{sec:conclusions} collects the conclusions and outlook. Detailed derivations and the numerical scheme are given in the supplementary material.

\section{The fractional Langevin equation}
\label{sec:FLE}

For simplicity, we first consider a non-relativistic heavy particle of mass $M$ undergoing one-dimensional Brownian motion in a thermal medium at temperature $T$. The stochastic dynamics is described by a coupled set of fractional Langevin equations~\cite{kobelev2000fractional,li2012spectral}:
\begin{align}
&\tau_{p}^{\,\beta-1}\,{}^{C}\!D^{\beta}_{0+}x(t) = \frac{p(t)}{M},
\label{eq:FLE_x_frac}\\[4pt]
&\tau_{p}^{\,\alpha-1}\,{}^{C}\!D^{\alpha}_{0+}p(t) = -\gamma\,p(t) + \xi(t),
\label{eq:FLE_p_frac}
\end{align}
where $p(t)$ and $x(t)$ denote the momentum and position of the particle at time $t$, and $^{C}D_{0+}^{\alpha}$, $^{C}D_{0+}^{\beta}$ are Caputo fractional derivatives of orders $\alpha$ and $\beta$, and for simplification we take $\alpha$ $\equiv$ $\beta$ . Since the Caputo operator ${}^{C}D^{\alpha}_{0+}$ on the left-hand side of the FLE does not carry the same dimension as the drag and stochastic-force terms on the right-hand side, dimensional consistency must be restored. To this end, a characteristic time $\tau_{p}$ is introduced on the left-hand side as an approximation, and the Caputo operator is multiplied by the prefactor $\tau_{p}^{\,\alpha-1}$. In the limit $\alpha\to 1$ the prefactor reduces to $\tau_{p}^{\,\alpha-1}\to 1$, so that the FLE reduces to the standard first-order Langevin equation independently of the specific value assigned to $\tau_{p}$. For $\alpha\neq 1$, a physically motivated value must be assigned to $\tau_{p}$; a natural candidate is the momentum-relaxation time of the HQ in the medium, given by the inverse drag coefficient,
\begin{align}
\tau_{p} \;\approx\; \frac{1}{\gamma} \;=\; \frac{M T}{\mathcal{D}} \;=\; \frac{M\, D_{s}}{T},
\label{eq:tau_p_def}
\end{align}
$\mathcal{D} = MT\gamma$ and $D_{s} = T/(M\gamma)$ are the momentum and spatial diffusion coefficients, respectively. The numerical magnitude of $\tau_{p}$ is inherited from lattice-QCD and phenomenological results. 
$(2{+}1)$-flavor lattice computations with dynamical fermions give $2\pi T D_{s}\approx 1$--$2$ for $T\simeq 195$--$293\,\mathrm{MeV}$, implying fast charm hydrodynamization~\cite{Altenkort:2023oms,Altenkort:2023eav}. Phenomenological extractions from $D$-meson $R_{AA}$ and $v_{2}$ at RHIC and the LHC are consistent: Boltzmann-transport fits yield a charm thermalization time $\tau_{\rm th}\sim 4$--$6\,\mathrm{fm}/c$ as $p\to 0$~\cite{Scardina:2017ipo}, and a Bayesian analysis of the same observables gives $2\pi T D_{s}\approx 1$--$3$ near $T_{c}$~\cite{Xu:2017obm}; see Ref.~\cite{He:2022ywp} for an overview. For $T\simeq 0.15$--$0.40\,\mathrm{GeV}$ and $M=1.3\,\mathrm{GeV}$, these inputs yield $\tau_{p}\simeq 0.4$--$3\,\mathrm{fm}/c$, the range used throughout the numerical analysis.

The stochastic force $\xi(t)$ is taken to be Gaussian white noise with vanishing mean,
\begin{align}
\langle \xi(t) \rangle = 0, \qquad
\langle \xi(t)\,\xi(t') \rangle = 2\mathcal{D}\,\delta(t-t'),
\label{eq:noise_corr}
\end{align}
and for $\alpha=1$ the drag and diffusion coefficients satisfy the standard fluctuation-dissipation relation $\gamma=\mathcal{D}/(MT)$~\cite{Moore:2004tg,Svetitsky:1987gq}. For $\alpha\neq 1$, the fractional derivative in the FLE modifies this relation. The Caputo fractional derivative~\cite{10.1111/j.1365-246X.1967.tb02303.x} is defined as
\begin{align}\label{Caputo}
^{C} D^\nu_{0+}u(t) = \frac{1}{\Gamma({n-\nu})}\int_0^t \frac{u^{(n)}(s)}{(t-s)^{1+\nu-n}}ds,
\end{align}
where $u^{(n)}$ is the $n$-th derivative of $u$; for superdiffusion ($1< \nu \leq 2$) the relevant form is obtained with $n=2$.

\section{Fractional fluctuation-dissipation theorem}
\label{sec:FFDT}

To determine the relation between $\mathcal{D}$ and $\gamma$ that ensures thermal equilibrium for the FLE, the stationary regime is considered. The Fourier transform of Eq.~\eqref{eq:FLE_p_frac} is taken, and the Caputo derivative maps to multiplication by $(i\omega)^{\alpha}$ in frequency space, giving

\begin{align}
\widetilde{p}(\omega) = \frac{\widetilde{\xi}(\omega)}{\gamma+\tau_{p}^{\,\alpha-1}(i\omega)^{\alpha}} \equiv H(\omega)\,\widetilde{\xi}(\omega),
\label{eq:p_Fourier}
\end{align}
with $H(\omega)=[\gamma+\tau_{p}^{\,\alpha-1}(i\omega)^{\alpha}]^{-1}$. The
momentum power spectral density $S_{p}$ and noise power spectral density $S_{\xi}$ are related by $S_{p}(\omega)=|H(\omega)|^{2}\,S_{\xi}(\omega)$ \cite{Khinchin1934,Wiener1930}. The noise power spectral density is the Fourier transform of this autocorrelation function of the thermal  noise, $S_{\xi}(\omega)=2\mathcal{D}$, which is independent of frequency; the flat spectrum is the defining property of white noise. Evaluating the equilibrium variance yields as 

\begin{align}
\langle p^{2}\rangle_{\mathrm{eq}} = \frac{2\mathcal{D}}{\alpha}\, \gamma^{\,1/\alpha-2}\, \tau_{p}^{-(\alpha-1)/\alpha}\, \frac{\sin\tfrac{\pi(\alpha-1)}{2}}{\sin\tfrac{\pi\alpha}{2}\,\sin\tfrac{\pi}{\alpha}}.
\label{eq:p2_final_tau}
\end{align}
Imposing the thermal-equilibrium condition $\langle p^{2}\rangle_{\mathrm{eq}}=MT$ and solving for $\mathcal{D}$ yields the F-FDT (as detailed in supplementary material)
\begin{align}
\mathcal{D}(\alpha) = K_{\alpha}\,MT\, \gamma^{\,2-1/\alpha}\, \tau_{p}^{\,(\alpha-1)/\alpha},
\label{eq:FDT_frac_tau}
\end{align}
where $K_{\alpha} \equiv \tfrac{\alpha}{2}\, \sin\tfrac{\pi\alpha}{2}\,\sin\tfrac{\pi}{\alpha}/\sin\tfrac{\pi(\alpha-1)}{2}$, valid for $1<\alpha\leq2$. In the limit $\alpha\to 1$, $K_{\alpha}\to 1$, and the standard fluctuation-dissipation relation $\mathcal{D}=\gamma MT$ is recovered. 

For the mean-square displacement, the same procedure applied to Eq.~\eqref{eq:FLE_x_frac} gives the position power spectral density
\begin{align}
S_{x}(\omega) = \frac{2\mathcal{D}}{M^{2}\,\tau_{p}^{\,2(\beta-1)}\,\omega^{2\beta}\, \bigl|\gamma+\tau_{p}^{\,\alpha-1}(i\omega)^{\alpha}\bigr|^{2}},
\label{eq:S_x}
\end{align}
and, through $\langle x^{2}(t)\rangle = \tfrac{2}{\pi}\int_{0}^{\infty} S_{x}(\omega)(1-\cos\omega t)\,d\omega$, the long-time behavior (as detailed in supplementary material)
\begin{align}
\langle x^{2}(t)\rangle \;\sim\; \frac{2\mathcal{D}}{M^{2}\,\gamma^{2}\,\tau_{p}^{\,2(\beta-1)}}\, \frac{t^{\,2\beta-1}}{\Gamma(2\beta)\,\sin\!\big(\pi(\beta-\tfrac12)\big)},
\label{eq:msd_asymptotic}
\end{align}
valid for  $t\gg\tau_{p}$, which exhibits the anomalous power law $\langle x^{2}(t)\rangle\propto t^{\,2\beta-1}$ characteristic of superdiffusive transport. In the limit $\alpha\to 1$ and $\beta\to 1$, this reduces to $\langle x^{2}(t)\rangle\to 2D_{s}t$,  in agreement with the standard Einstein relation employed in the conventional Langevin description of HQ transport in the QGP~\cite{Moore:2004tg,Svetitsky:1987gq}. We now proceed to solve the Langevin equation, Eqs.~\eqref{eq:FLE_x_frac},~\eqref{eq:FLE_p_frac}, for HQs in the relativistic regime so, $M\to E$ with $E=\sqrt{p^{2}+M^{2}}$, and the fractional generalization reads
\begin{align}
&\tau_{p}^{\,\beta-1}\,{}^{C}\!D^{\beta}_{0+}x(t) = \frac{p(t)}{E(t)},
\label{eq:FLE_x_frac_rel}\\[4pt]
&\tau_{p}^{\,\alpha-1}\,{}^{C}\!D^{\alpha}_{0+}p(t) = -\gamma\,p(t) + \xi(t).
\label{eq:FLE_p_frac_rel}
\end{align}
We impose the F-FDT in the relativistic limit, namely
\begin{align}
\mathcal{D}(\alpha) = K_{\alpha}\,ET\, \gamma^{\,2-1/\alpha}\, \tau_{p}^{\,(\alpha-1)/\alpha},
\label{eq:FDT_frac_tau_rel}
\end{align}
the FLE for the HQ momentum and position is solved numerically using a three-level central-difference L2 scheme for the Caputo derivative; the numerical scheme  on the HQ dynamics are detailed in supplementary material. 

\section{Results}
\label{sec:results}
All results are obtained for a charm quark of mass $M=1.3~\mathrm{GeV}$, diffusion coefficient,  $\mathcal{D}=0.2~\mathrm{GeV}^{2}/\mathrm{fm}$, drag coefficient $\gamma$ fixed through the F-FDT of Eq.~\eqref{eq:FDT_frac_tau_rel} and the characteristic time set to $\tau_{p}=2~\mathrm{fm}/c$. 


We first consider a charm quark initialized at rest and study the evolution of $\langle p^{2}(t)\rangle$, which tests the fractional fluctuation--dissipation
relation through its approach to thermal equilibrium.  We then calculate the   final transverse-momentum distribution $dN/dp_{T}$, the average energy loss $\langle \Delta E\rangle$, the longitudinal momentum loss $\langle \Delta p_{\parallel}\rangle$, and the transverse momentum broadening $\langle p_{\perp}^{2}\rangle$, which characterize energy dissipation, loss of directed momentum, and broadening in the medium, respectively.  We calculate the nuclear modification factor $R_{AA}(p_{T})$  to study the suppression and medium-induced modification of the charm quark distribution. A path-length estimate of the elliptic flow, $v_{2}(p_{T})$, is also calculated from the difference between in-plane and out-of-plane suppression to show the sensitivity of the fractional dynamics.

\begin{figure}[t]
\centering
\includegraphics[scale = .35]{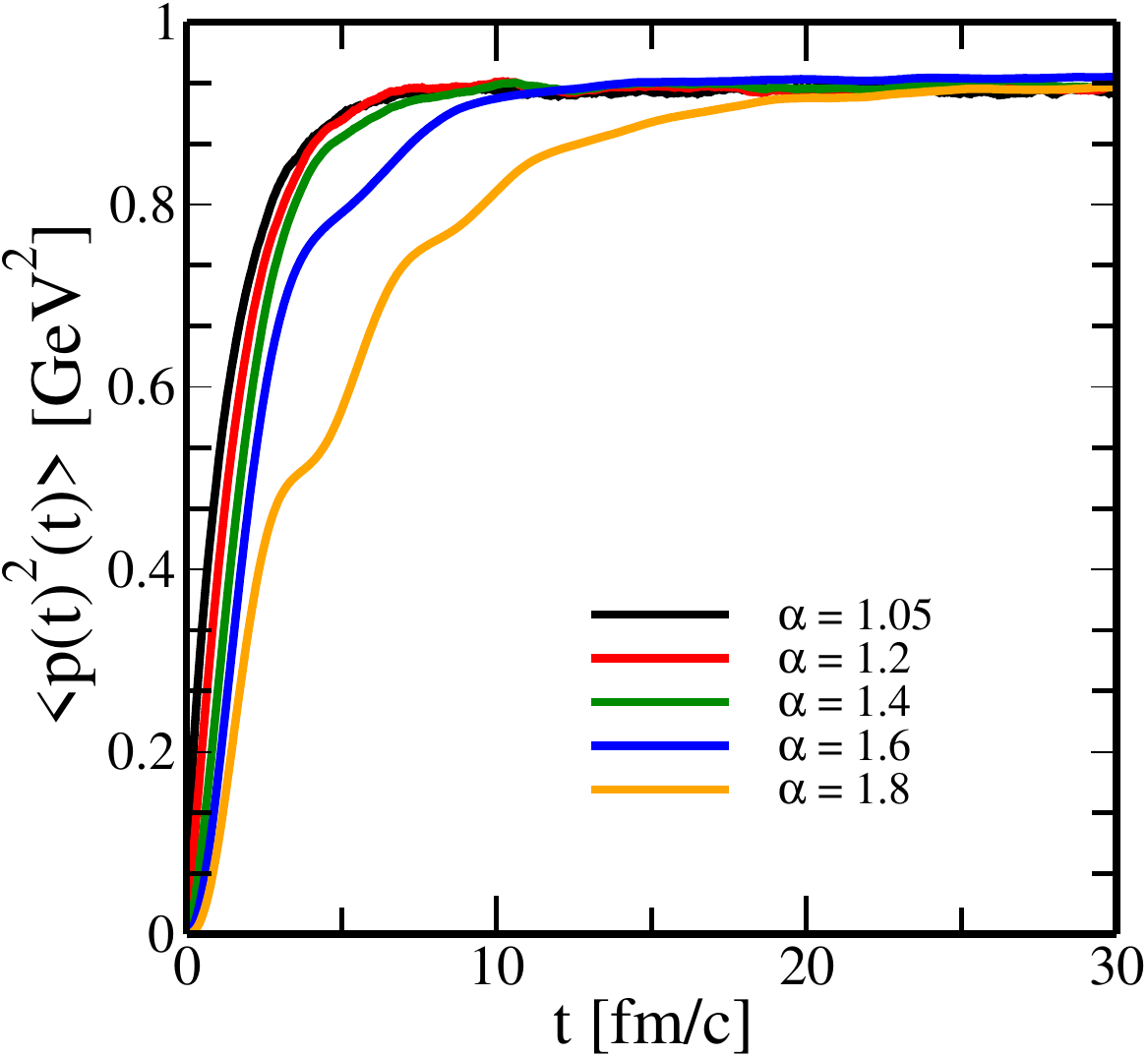}
\caption{Time evolution of $\langle p^{2}(t)\rangle$ for  charm quarks at several values of $\alpha$, with $\mathcal{D}=0.2~\mathrm{GeV}^{2}/\mathrm{fm}$, $\tau_{p}=2~\mathrm{fm}/c$, and $T=0.25~\mathrm{GeV}$.}
\label{fig:p2_BC}
\end{figure}

The time evolution of the HQs $\langle p^{2}(t)\rangle$ in the QGP medium is shown in Fig.~\ref{fig:p2_BC} for charm quarks. The curves correspond to five representative values of the fractional order, $\alpha=1.05, 1.2$, $1.4$, $1.6$, and $1.8$, with the noise strength fixed by the F-FDT. The charm quarks are initialised with $p_{x}(t_{0})=p_{y}(t_{0})=0$. At early times $\langle p^{2}(t)\rangle$ grows rapidly, and at late times it approaches a plateau corresponding to thermal equilibrium with the QGP medium. The approach to thermalization is controlled by $\alpha$: smaller values of $\alpha$ produce steeper early-time growth and earlier saturation, whereas larger values lead to slower broadening and a more gradual approach to the plateau. The curves for a given species converge to the same asymptotic value for all values of $\alpha$.  In the non-relativistic limit $E\to M$, the equilibrium relation reduces to the standard equipartition result $\langle p^{2}\rangle_{\rm eq}=2MT$ \cite{Chen:2023pgx}.

\begin{figure}[t]
\centering
\includegraphics[scale = .35]{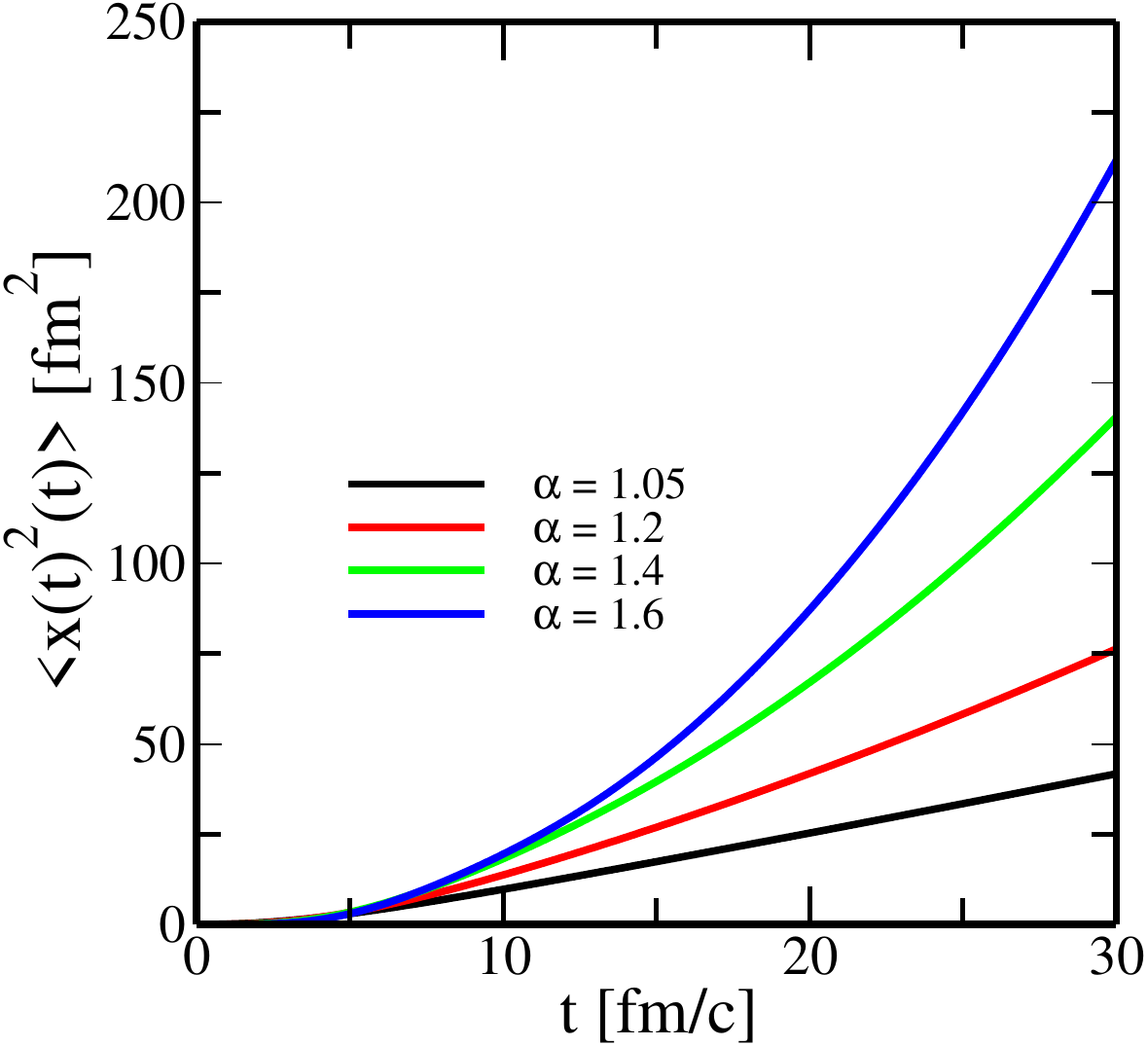}
\caption{Time evolution of $\langle x^{2}(t)\rangle$ of charm  quarks for several values of $\alpha$, with $\mathcal{D} = 0.2~\mathrm{GeV}^{2}/\mathrm{fm}$, $\tau_p =2$~fm/$c$, and $T = 0.25$~GeV.}
\label{fig:x2_BC}
\end{figure}

The corresponding time evolution of $\langle x^{2}(t)\rangle$ is shown in Fig.~\ref{fig:x2_BC} for charm quarks, for $\alpha=1.05$, $1.2$, $1.4$, and $1.6$, with the remaining simulation parameters kept the same as in Fig.~\ref{fig:p2_BC}. In contrast to $\langle p^{2}(t)\rangle$, $\langle x^{2}(t)\rangle$ does not saturate over the simulated time and continues to grow monotonically. The curves cross at intermediate times, and at later times a clear ordering with $\alpha$ emerges: larger values of $\alpha$ yield a faster growth of $\langle x^{2}(t)\rangle$.


\begin{figure}[t]
	\centering
	\includegraphics[scale = .3]{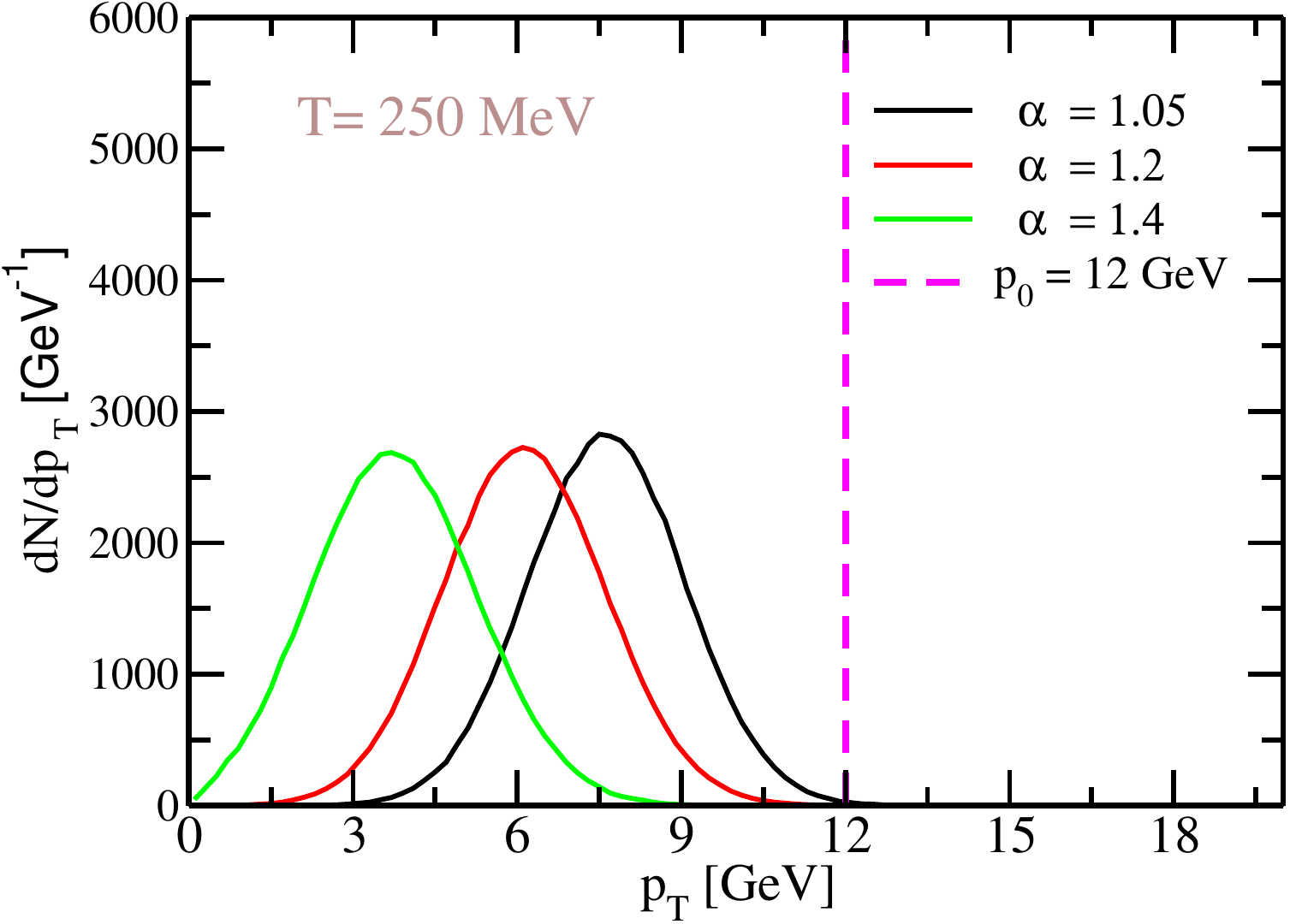}\hfill
	\includegraphics[scale = .3]{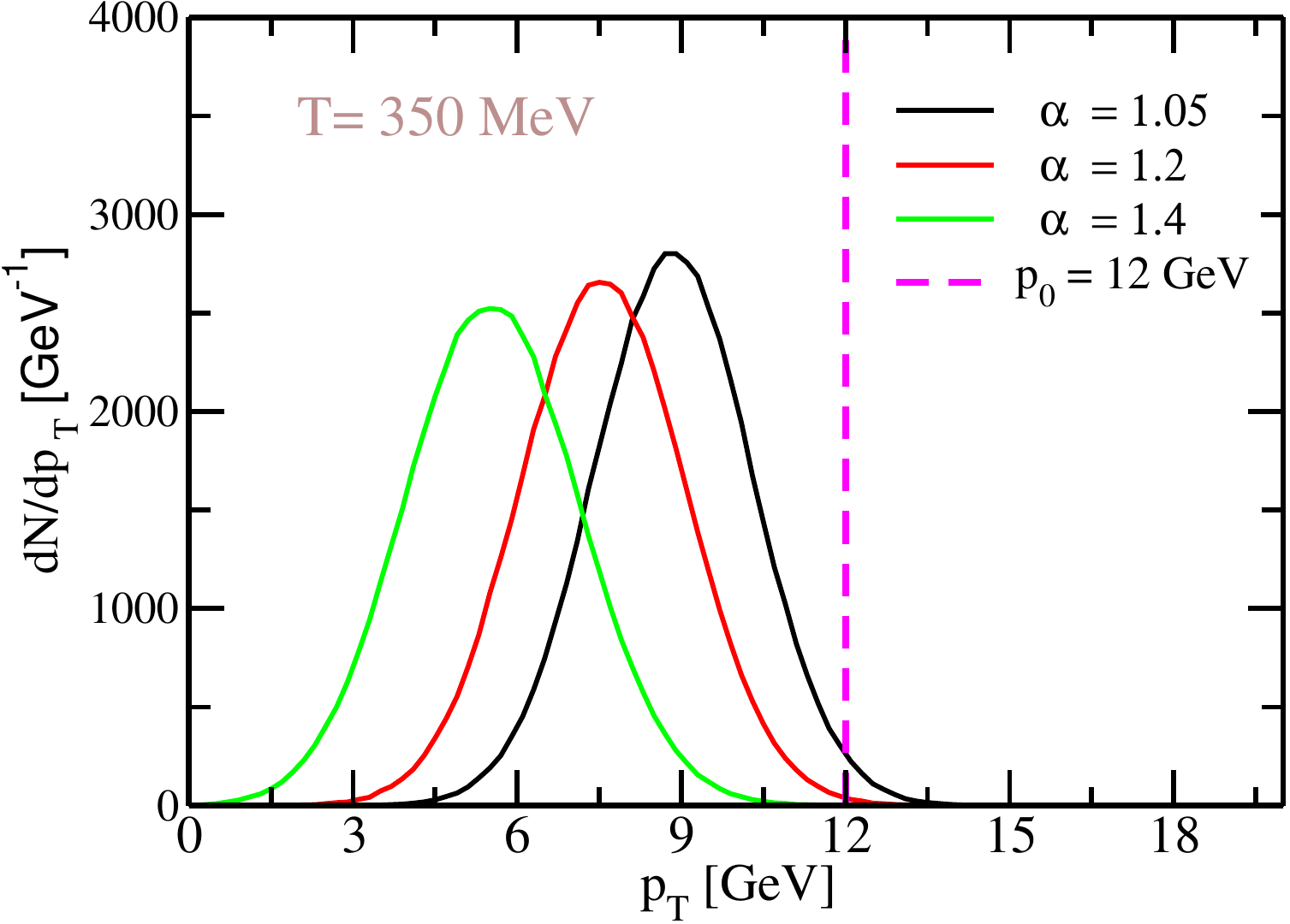}
	\caption{{Evolution of the charm-quark momentum distribution at $t=5$~fm/$c$ as a function of $p_T$, for various $\alpha$ and two temperatures [$T=0.25$~GeV (above) and $T=0.35$~GeV (below)], for an initial delta distribution centered at $p(t_0)=12$~GeV.}}
	\label{momentum_distribution}
\end{figure}

The evolution of $dN/dp_{T}$ of the charm quark is computed at $T=0.25$~GeV and $T=0.35$~GeV in Fig.~\ref{momentum_distribution}, for $\alpha=1.05$, $1.2$,  and $1.4$, at a final evolution time $t=5$~fm/$c$, starting from a delta-like distribution centered at $p_{0}=12$~GeV. At $T=0.25$~GeV, $\alpha=1.05$ reproduces the standard Langevin result reported in Refs.~\cite{Das:2013kea} for the same input parameters. As $\alpha$ increases from $1.05$ to $1.4$, the distribution broadens, and the peak shifts away from the initial momentum, with the most pronounced spread and displacement at $\alpha=1.4$. At $T=0.35$~GeV, the same ordering continues; at fixed $\alpha$, it broadens more strongly. The total area under each curve is conserved.


We further initialize charm quarks  with $p_{T}=100~\mathrm{GeV}$  and set the medium temperature, $T=0.4~\mathrm{GeV}$. The fractional orders are $\alpha=1.001$, $1.2$, $1.4$, $1.6$, and $1.8$, with $\alpha=1.001$ serving as the near-Markovian reference.

The observables are defined relative to the initial momentum direction $\widehat{\boldsymbol n}_0=\boldsymbol p_0/|\boldsymbol p_0|$. For each charm quark,
\begin{align}
p_{\parallel}(t)
  &= \boldsymbol p(t)\!\cdot\!\widehat{\boldsymbol n}_0,
\nonumber\\
p_{\perp}^{2}(t)
  &= |\boldsymbol p(t)|^{2}-p_{\parallel}^{2}(t),
\nonumber\\
\Delta p_{\parallel}(t)
  &= |\boldsymbol p_0|-p_{\parallel}(t),
\nonumber\\
\Delta E(t)
  &= \sqrt{|\boldsymbol p_0|^{2}+M^{2}}
   -\sqrt{|\boldsymbol p(t)|^{2}+M^{2}}.
\label{eq:loss_observables}
\end{align}
The quantities displayed are the corresponding ensemble averages, $\langle\Delta E\rangle$, $\langle\Delta p_{\parallel}\rangle$, and $\langle p_{\perp}^{2}\rangle$. Note that $p_{\parallel}$ and $p_{\perp}$ are resolved with respect to $\widehat{\boldsymbol n}_0$.
\begin{figure*}[!tb]
\centering
\includegraphics[width=0.32\textwidth]{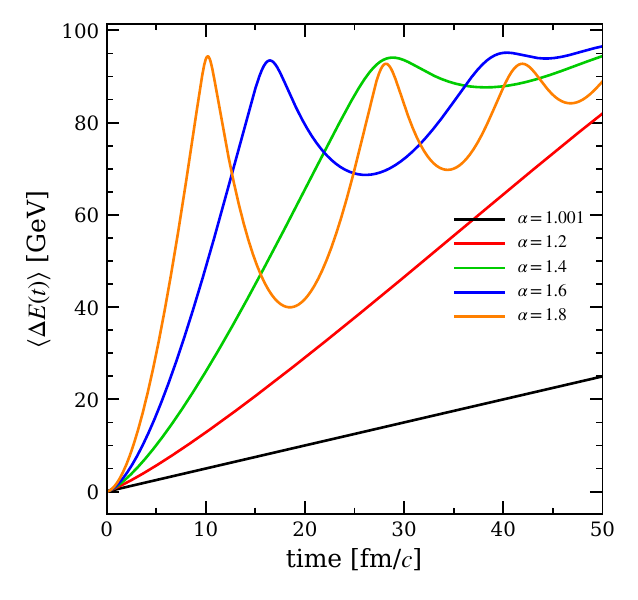}\hfill
\includegraphics[width=0.32\textwidth]{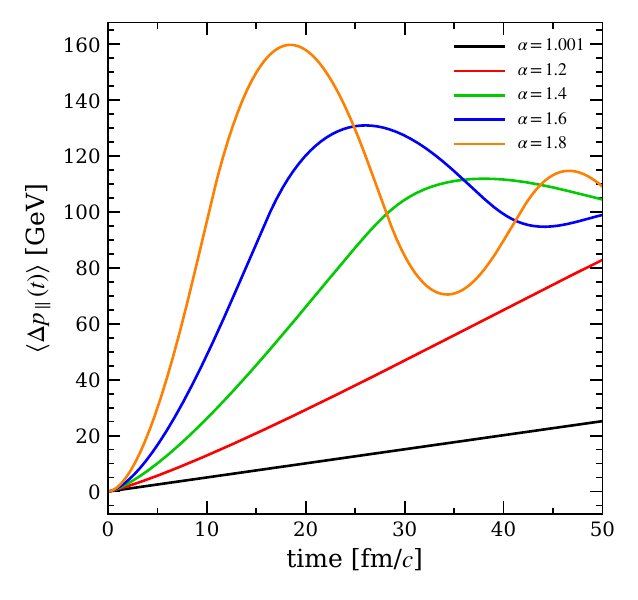}\hfill
\includegraphics[width=0.32\textwidth]{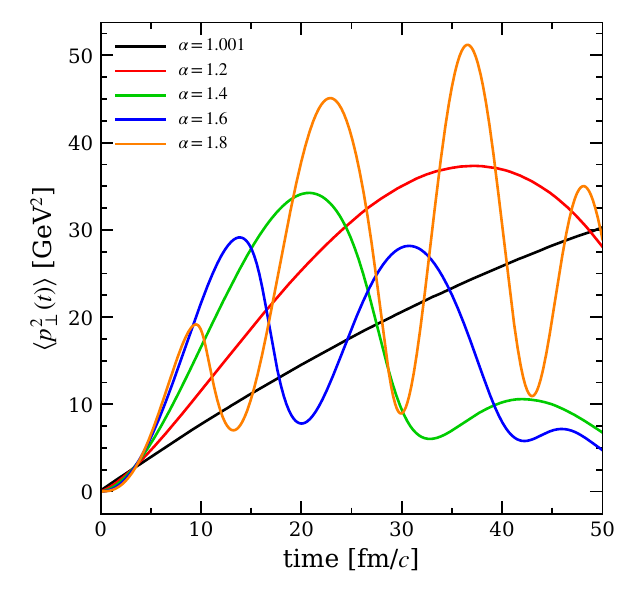}
\caption{Time evolution of the charm quarks. From left to right: mean energy loss $\langle\Delta E(t)\rangle$, mean longitudinal momentum loss $\langle\Delta p_{\parallel}(t)\rangle$, and transverse momentum broadening $\langle p_{\perp}^{2}(t)\rangle$.}
\label{fig:time_observables_constant_DT}
\end{figure*}

The left panel of Fig.~\ref{fig:time_observables_constant_DT} shows the evolution of the mean energy loss. The near-Markovian case, $\alpha=1.001$, exhibits a slow and smooth increase of the energy loss throughout the evolution, indicating that the heavy quark retains a substantial fraction of its initial energy even at late times. Increasing the fractional order leads to a much faster degradation of the initial energy. For $\alpha\gtrsim1.4$, however, the relaxation ceases to be monotonic and pronounced maxima and minima develop. These oscillations become particularly strong for $\alpha=1.8$, showing that the larger fractional orders do not simply enhance the energy loss uniformly, but instead introduce a strong underdamped oscillatory relaxation response in which the energy can partially recover after a period of rapid degradation. Thus, the larger values of $\alpha$ substantially accelerate the departure from the initial high-energy state, while simultaneously producing increasingly nonmonotonic relaxation.

The middle panel of Fig.~\ref{fig:time_observables_constant_DT} displays the corresponding longitudinal momentum loss. The $\alpha=1.001$ result again evolves gradually, whereas increasing $\alpha$ causes the initial directed momentum to relax much more rapidly. For the larger fractional orders, the longitudinal momentum loss approaches the scale set by the initial momentum already within the evolution time considered. Moreover, for some values of $\alpha$ it temporarily exceeds this scale, which corresponds to a reversal of the ensemble-averaged longitudinal momentum rather than an increase of the momentum magnitude. The subsequent decrease of the longitudinal loss therefore reflects the oscillatory return of the mean momentum after this reversal. The increasingly pronounced overshoot with increasing $\alpha$ provides a direct manifestation of the nonmonotonic, underdamped relaxation characteristic of the fractional dynamics.

The right panel of Fig.~\ref{fig:time_observables_constant_DT} shows the transverse momentum broadening. Starting from zero, the transverse fluctuations are generated by the stochastic interaction with the medium and initially increase for all fractional orders. The $\alpha=1.001$ case develops smoothly over the full evolution, whereas larger values of $\alpha$ produce increasingly pronounced maxima and minima. For the intermediate fractional orders, particularly $\alpha=1.4$ and $\alpha=1.6$, the transverse broadening decreases substantially after its initial maximum and moves toward a considerably smaller momentum scale at late times. This behavior is consistent with the progressive loss of the initial motion and the approach toward a thermal momentum distribution. In contrast, the $\alpha=1.8$ result remains strongly oscillatory, indicating that the system repeatedly departs from and returns toward the low-momentum regime instead of approaching it monotonically.
 

\begin{figure*}[!tb]
\centering
\includegraphics[width=0.32\textwidth]{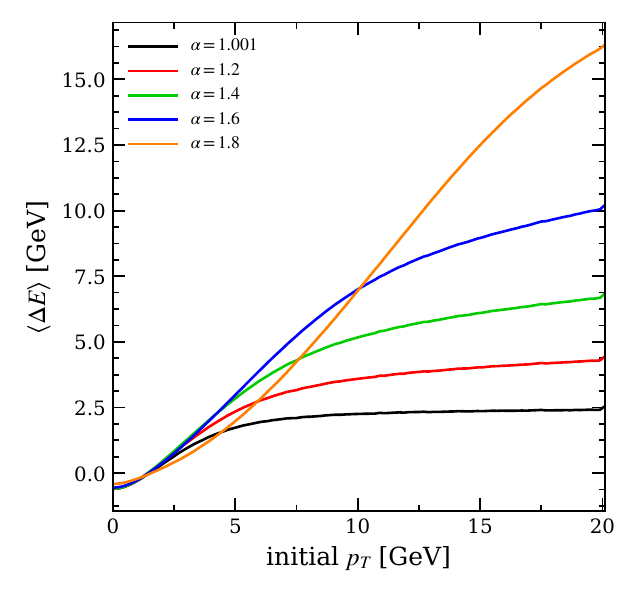}\hfill
\includegraphics[width=0.32\textwidth]{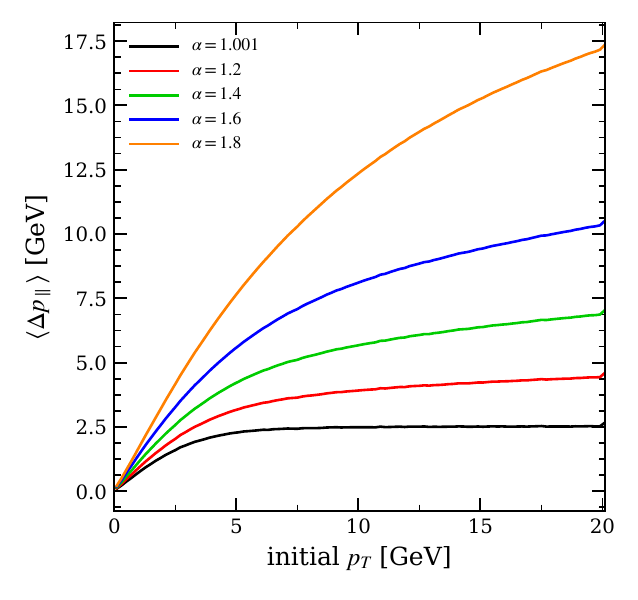}\hfill
\includegraphics[width=0.32\textwidth]{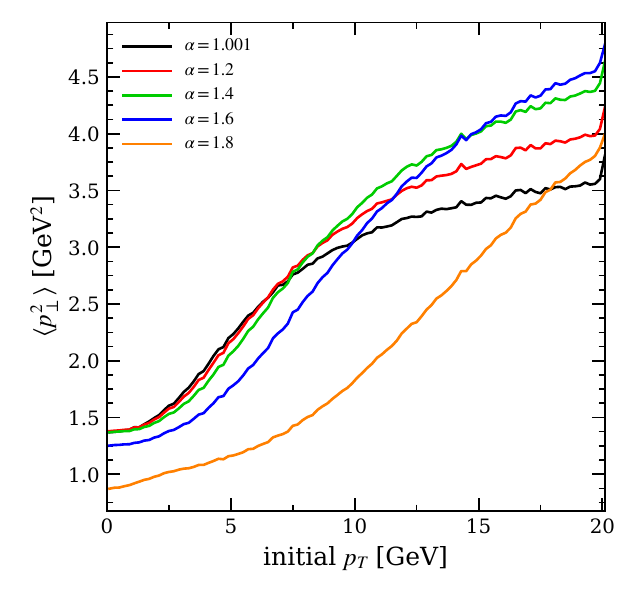}
\caption{Initial-momentum dependence of the charm-quark quantities. From left to right: $\langle\Delta E\rangle$, $\langle\Delta p_{\parallel}\rangle$, and $\langle p_{\perp}^{2}\rangle$.}
\label{fig:pt_observables_constant_DT}
\end{figure*}

To calculate the dependence on the initial heavy-quark momentum, the system is evolved up to $t=5~\mathrm{fm}/c$ for $D=0.2~\mathrm{GeV}^{2}/\mathrm{fm}$ and $T=0.4~\mathrm{GeV}$. The left panel of Fig.~\ref{fig:pt_observables_constant_DT} shows the mean energy loss as a function of the initial transverse momentum. For the near-Markovian case, $\alpha=1.001$, the energy loss depends only weakly on the initial momentum and remains comparatively small over the entire $p_T$ range considered. As the fractional order increases, the energy degradation becomes progressively stronger, and the separation among the different $\alpha$ curves becomes particularly pronounced at large initial momentum. For the largest fractional orders, a substantial fraction of the initial energy can already be dissipated within $5~\mathrm{fm}/c$. This behavior demonstrates that the fractional dynamics significantly enhances the relaxation of energetic heavy quarks and that the sensitivity to the fractional order becomes increasingly important toward the high-$p_T$ region.

The middle panel of Fig.~\ref{fig:pt_observables_constant_DT} shows the corresponding longitudinal momentum loss. Its dependence on both the initial momentum and the fractional order closely follows that of the energy loss. The near-Markovian case produces only a moderate reduction of the momentum component along the initial propagation direction, whereas increasing $\alpha$ leads to a substantially stronger longitudinal degradation. At large initial $p_T$, the higher-$\alpha$ cases remove a considerable fraction of the initial directed momentum within the same evolution time. The close correspondence between the energy and longitudinal-momentum losses indicates that the enhanced energy degradation is predominantly associated with the relaxation of the momentum component along the initial direction. 
The right panel of Fig.~\ref{fig:pt_observables_constant_DT} presents the transverse momentum broadening. In contrast to the energy and longitudinal-momentum losses, its dependence on the fractional order is not monotonic. The transverse broadening generally increases with the initial momentum, indicating stronger transverse fluctuations for more energetic heavy quarks. However, the different $\alpha$ curves cross as $p_T$ increases, and the ordering observed at low momentum is not preserved in the high-$p_T$ region. 



The nuclear suppression factor $R_{AA}(p_{T})$ for the HQs is calculated to analyze the impact of the fractional dynamics on the experimental observable. It is defined as~\cite{Moore:2004tg}
\begin{align}\label{raa_e}
	R_{AA}(p_{T}) = \frac{f_{t}(p_{T})}{f_{t_0}(p_{T})},
\end{align}
where $f_{t}(p)$ is the charm-quark momentum spectrum at the evolution time, $t$ and $f_{t_0}(p)$ is the initial momentum distribution. The initial momentum spectra, $f_{t_0}(p)$ is taken through a weight obtained from a fit to the fixed-order plus next-to-leading-logarithmic (FONLL) calculation, which reproduces the measured $D$-meson spectra measured in proton--proton collisions through fragmentation~\cite{cacciari2005qcd,Cacciari:2012ny}. The fit function is therefore a parametrization of the FONLL spectrum, \begin{equation}
	dN/d^{2}p_{T} = x_{0}/(x_{1}+p_{T})^{x_{2}}\end{equation} and $x_{0}=6.36\times 10^{8}$, $x_{1}=9.0$, $x_{2}=10.28$, where $p_{T}=\sqrt{p_{x}^{2}+p_{y}^{2}}$. A value $R_{AA}(p_{T})\neq 1$ indicates that the charm quarks have interacted with the medium and that their spectrum has been modified relative to the initial distribution.

\begin{figure}[t]
	\centering
	\includegraphics[scale = .3]{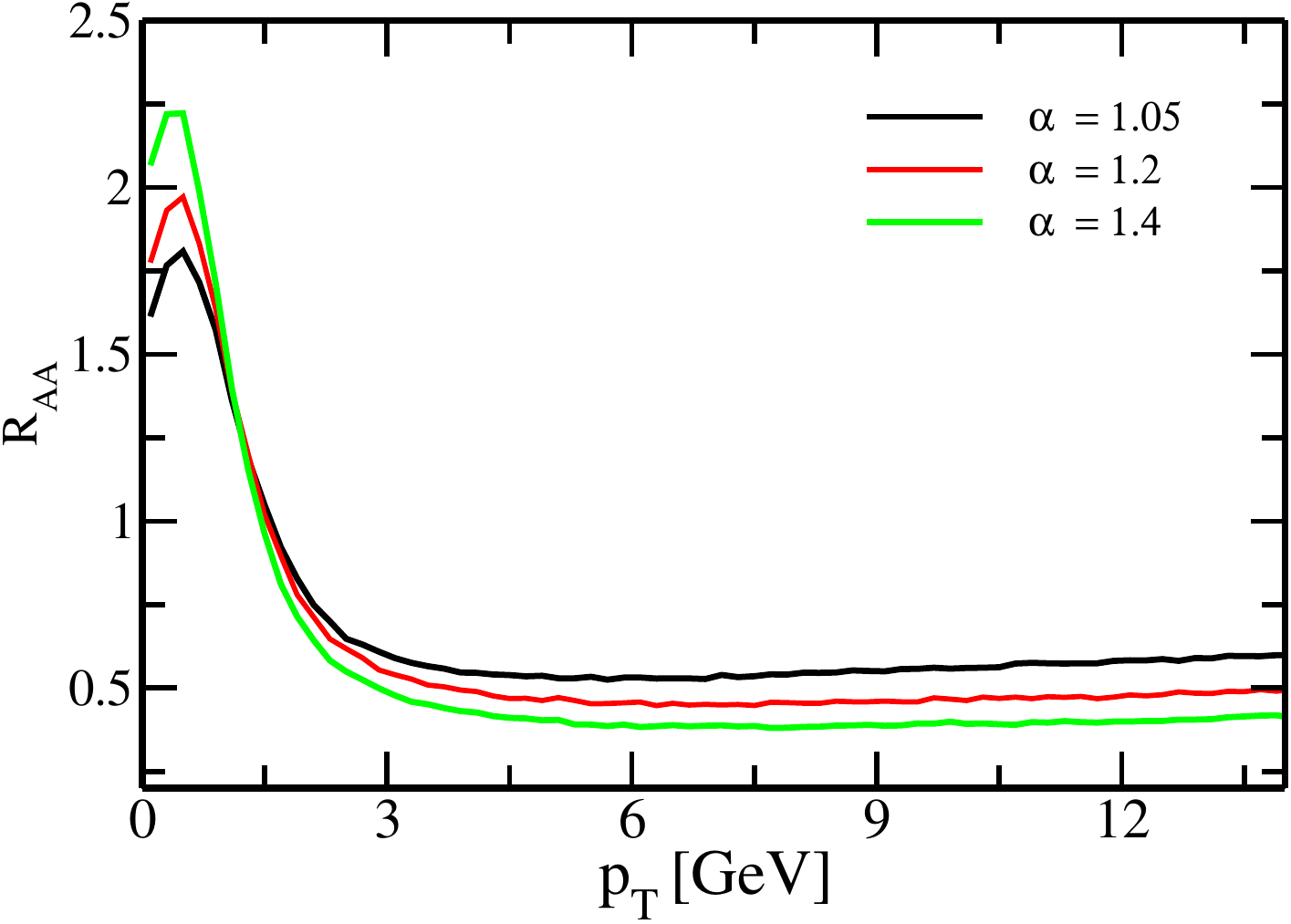}\hfill
	\includegraphics[scale = .3]{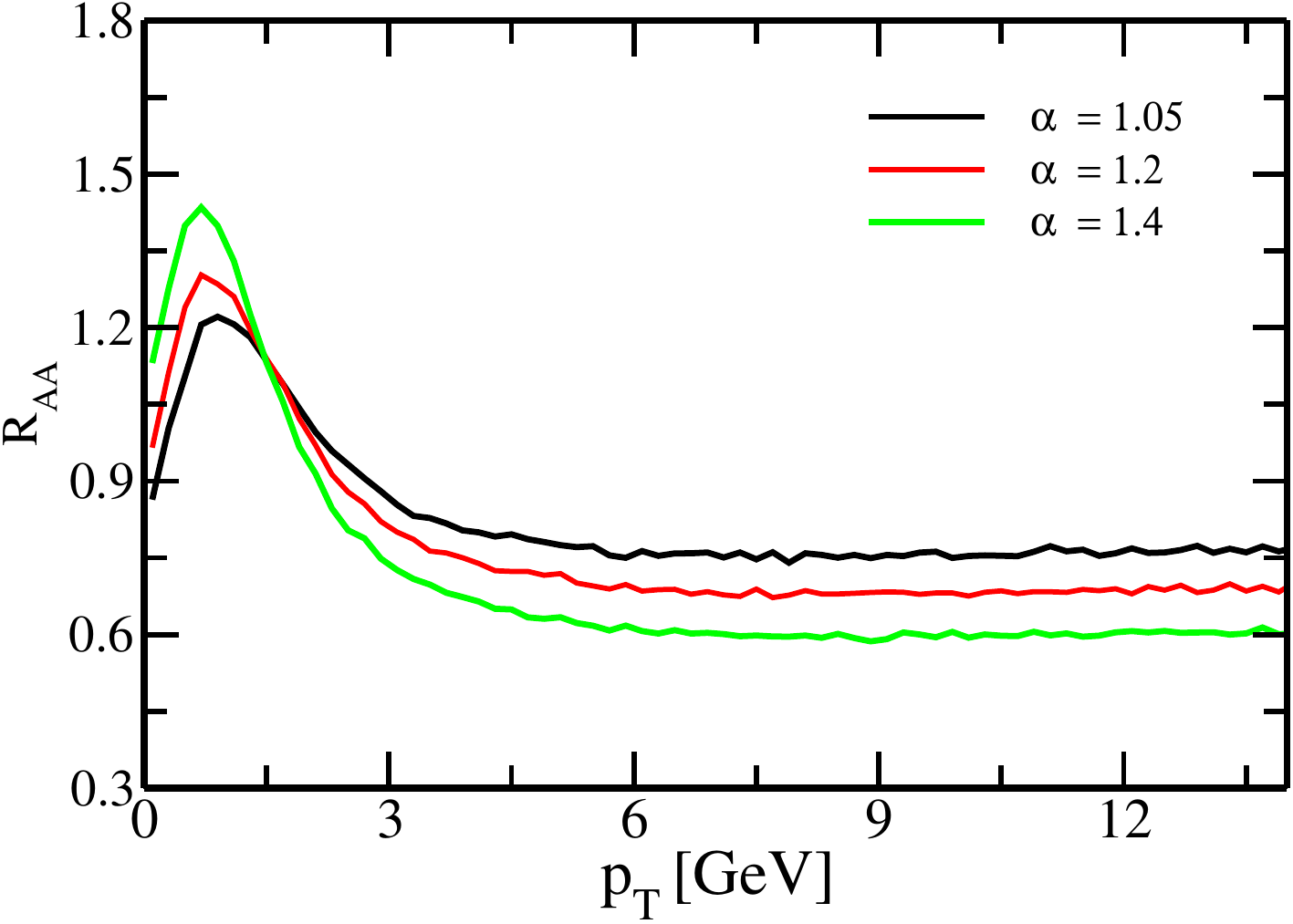}
	\caption{{The $R_{AA}$ as a function of $p_T$, with $t=2$~fm/$c$, at two temperatures [$T=0.25$~GeV (above) and $T=0.5$~GeV (below)] for three values of $\alpha$.}}
	\label{raa}
\end{figure}

Figure~\ref{raa} shows $R_{AA}$ as a function of $p_{T}$, computed from Eq.~\eqref{raa_e} within the FLE, at $T=0.25$~GeV and $T=0.5$~GeV for $\alpha=1.05$, $1.2$, and $1.4$. The $\alpha\to 1$ limit, represented in the figure by $\alpha=1.05$, reproduces the standard Langevin result of Refs.~\cite{Cao:2013ita,scardina2017estimating} for the same input parameters and serves as a consistency check on the present framework. At $T=0.25$~GeV, $R_{AA}$ falls below unity at high $p_{T}$; as $\alpha$ increases from $1.2$ to $1.4$, the suppression at high $p_{T}$ deepens monotonically and the enhancement above unity at low $p_{T}$ becomes more pronounced, consistent with the conservation of the total charm-quark number throughout the evolution.

Before comparing the two path lengths, the overall suppression is first matched at $t=5~\mathrm{fm}/c$. To this end we write the transport coefficient as $\mathcal{D}(\alpha)=N_D(\alpha)\,\mathcal{D}_0$, with
$\mathcal{D}_0=0.2~\mathrm{GeV}^2/\mathrm{fm}$ and $N_D(\alpha)$ a dimensionless multiplier that rescales the transport strength at fixed
$\alpha$. The near-Markovian calculation, $\alpha=1.001$ with $N_D=1$,
defines the reference $R_{AA}$. For each $\alpha=1.2$, $1.4$, $1.6$, and
$1.8$, $N_D(\alpha)$ is then the only quantity tuned, and is adjusted so
that $R_{AA}$ integrated over $4\le p_T\le 10~\mathrm{GeV}$ approximately
matches this reference. 
The fitted coefficient is then frozen for that $\alpha$ and used unchanged to calculate $R_{AA}$ at the shorter in-plane time $t_{\rm in}=4~\mathrm{fm}/c$ and the longer out-of-plane time $t_{\rm out}=6~\mathrm{fm}/c$. 
\begin{figure}[t]
	\centering
	\includegraphics[width=\linewidth]{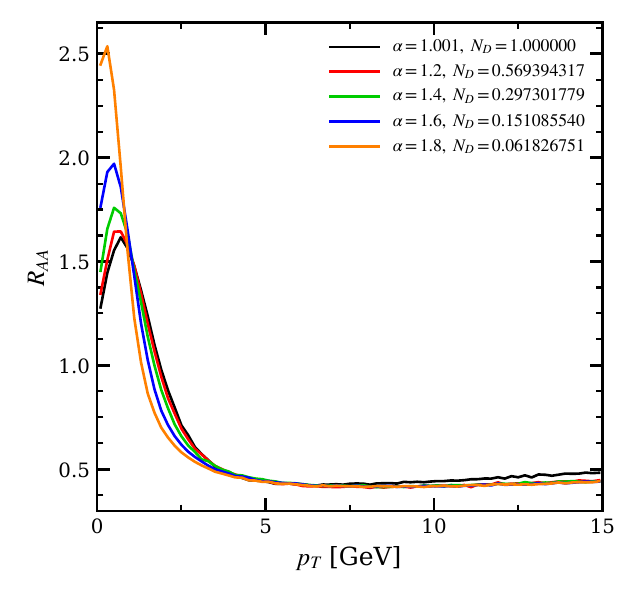}
	\caption{Charm-quark $R_{AA}$ at $t=5~\mathrm{fm}/c$. The $\alpha=1.001$ result with $N_D=1$.}
	\label{fig:normalized_raa_5fm}
\end{figure}
\begin{figure}[t]
	\centering
	\includegraphics[width=\linewidth]{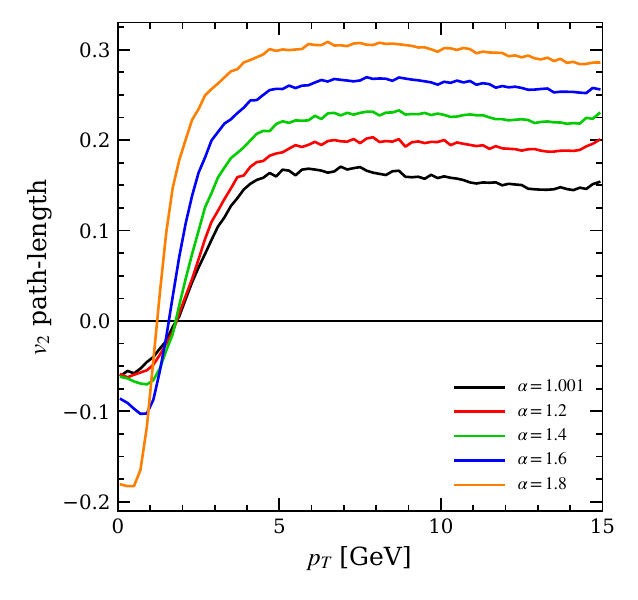}
	\caption{Path-length estimate of the charm quark elliptic-flow coefficient $v_2$, obtained from Eq.~\eqref{eq:v2_path_length} with $t_{\rm in}=4~\mathrm{fm}/c$ and $t_{\rm out}=6~\mathrm{fm}/c$, as a function of $p_T$ for $\alpha=1.001$, $1.2$, $1.4$, $1.6$, and $1.8$.}
	\label{fig:v2_path_length}
\end{figure}

Figure~\ref{fig:normalized_raa_5fm} shows the matched spectra that define this calibration. Retaining only the subsequent path-length asymmetry, a quark emitted in plane traverses less medium and is less suppressed than one emitted out of plane. 

The  $v_{2}$  is calculated  as 
\begin{align}
	v_{2}^{\rm path}(p_{T}) \;\approx\; \frac{1}{2}
	\frac{R_{AA}^{\rm in}(p_{T})-R_{AA}^{\rm out}(p_{T})}
	{R_{AA}^{\rm in}(p_{T})+R_{AA}^{\rm out}(p_{T})},
	\label{eq:v2_path_length}
\end{align}
where $R_{AA}^{\rm in}$ and $R_{AA}^{\rm out}$ are evaluated from Eq.~\eqref{raa_e} at $t_{\rm in}$ and $t_{\rm out}$, respectively. The temperature is $T=0.4~\mathrm{GeV}$, and the tuned $\mathcal{D}(\alpha)$ is constant during both evolutions. 
Figure~\ref{fig:v2_path_length} presents the  $v_2^{\rm path}$. It is negative below $p_T\simeq2$~GeV, changes sign near $2$~GeV, and then rises rapidly before forming a broad maximum at intermediate momentum. Around $p_T=6$--$8$~GeV, $v_2^{\rm path}$ increases  for $\alpha=1.8$ as compare to $\alpha=1.001$ at higher $p_T$. Because the overall suppression was matched first, this separation probes the different path-length sensitivity of the fractional dynamics rather than a difference in its baseline strength.

\section{Conclusions and outlook}
\label{sec:conclusions}

A framework for HQ transport in the QGP has been developed based on the fractional Langevin equation in the Caputo derivative formulation. A fractional fluctuation-dissipation theorem (F-FDT) has been derived directly from the FLE, fixing the relation between the drag coefficient $\gamma$ and the diffusion coefficient $\mathcal{D}$ for arbitrary $\alpha$ and reducing to the standard Einstein relation in the $\alpha\to 1$ limit. Within this framework, $\langle p^{2}(t)\rangle$ converges at large times to a common $\alpha$-independent value, while $\langle x^{2}(t)\rangle$ exhibits superdiffusive growth controlled by $\alpha$. The near-Markovian case $\alpha=1.05$ reproduces the standard Langevin result reported in the literature, providing a consistency check on the framework. As $\alpha$ increases, $R_{AA}$ at high $p_T$ shows progressively stronger suppression, and the momentum distribution $dN/dp_{T}$ broadens around the initial momentum with the peak shifting away from $p_{0}$, most pronounced at the largest $\alpha$. 

In the energetic regime, the fractional order controls the rate at which a fast charm quark is degraded. The mean energy loss and the longitudinal momentum loss grow markedly with $\alpha$ at fixed initial momentum, and their separation widens toward high $p_T$, so that energetic charm quarks are the most sensitive to the fractional dynamics. The relaxation becomes nonmonotonic for $\alpha\gtrsim1.4$, the Caputo kernel producing an oscillatory approach to equilibrium. The transverse momentum broadening inherits the same oscillatory structure, and its ordering with $\alpha$ is not uniform. Carried into a path-length estimate of the elliptic flow, this stronger energy-loss response also enhances the geometric momentum anisotropy, with $v_2$ at intermediate $p_T$ rising by roughly a factor of two between the near-Markovian case and $\alpha=1.8$.

Natural extensions include time-correlated noise and an expanding QGP background. The $v_2$ reported here isolates the path-length dependence alone; while $R_{AA}$ develops rapidly at the early stages of the QGP, $v_2$ builds up later, once the bulk medium has established collective flow, and a quantitative treatment therefore requires realistic initial conditions for the fireball geometry and expansion together with direct comparison to experimental data. A combined analysis of $R_{AA}$ and $v_2$ within the fractional framework may help constrain the transport coefficient via the fractional parameters. It would also be  interesting to study how the fractional dynamics modify the momentum broadening, i.e.  $\hat{q}$, and thereby affect the gluon emission responsible for radiative energy loss. These are left for future work.

\section*{Acknowledgements}
J.~P. acknowledges Dr.~Santosh Kumar Das for suggestions, and Aditi Tomar for insightful discussions and encouragement. This work was supported by the National Natural Science Foundation of China (NSFC) under Grant Nos.~12422508, 12375124, 12575146, 12225503  and the Science and Technology Commission of Shanghai Municipality under Grant No.~23JC1402700. Y.~S. thanks the sponsorship from Yangyang Development Fund.

\printbibliography

@article{Sambataro:2025obe,
    author = "Sambataro, Maria Lucia and Minissale, Vincenzo and Plumari, Salvatore and Greco, Vincenzo",
    title = "{Assessing lattice QCD charm space diffusion coefficient and thermalization time by mean of D meson observables at LHC}",
    eprint = "2508.01024",
    archivePrefix = "arXiv",
    primaryClass = "hep-ph",
    doi = "10.1016/j.physletb.2025.140049",
    journal = "Phys. Lett. B",
    volume = "872",
    pages = "140049",
    year = "2026"
}

@article{Grishmanovskii:2023gog,
    author = "Grishmanovskii, Ilia and Soloveva, Olga and Song, Taesoo and Greiner, Carsten and Bratkovskaya, Elena",
    title = "{Inelastic and elastic parton scattering in the strongly interacting quark-gluon plasma}",
    eprint = "2308.03105",
    archivePrefix = "arXiv",
    primaryClass = "hep-ph",
    doi = "10.1103/PhysRevC.109.024911",
    journal = "Phys. Rev. C",
    volume = "109",
    number = "2",
    pages = "024911",
    year = "2024"
}

@article{Oei:2024pva,
    author = "Oei, Naomi and Krenz, Nadja and van Hees, Hendrik and Greiner, Carsten and Torres-Rincon, Juan M.",
    title = "{Formation, dissociation, and regeneration of charmonia within microscopic Langevin simulations}",
    eprint = "2410.19619",
    archivePrefix = "arXiv",
    primaryClass = "hep-ph",
    doi = "10.1103/PhysRevD.111.074012",
    journal = "Phys. Rev. D",
    volume = "111",
    number = "7",
    pages = "074012",
    year = "2025"
}

@article{Grishmanovskii:2025mnc,
    author = "Grishmanovskii, Ilia and Song, Taesoo and Greiner, Carsten and Bratkovskaya, Elena",
    title = "{Transport coefficients of heavy quarks by elastic and radiative scatterings in the strongly interacting quark-gluon plasma}",
    eprint = "2503.22311",
    archivePrefix = "arXiv",
    primaryClass = "hep-ph",
    doi = "10.1103/rb24-dg58",
    journal = "Phys. Rev. D",
    volume = "112",
    number = "1",
    pages = "014042",
    year = "2025"
}

@article{Chen:2023pgx,
    author = "Chen, Weiguo and Greiner, Carsten and Xu, Zhe",
    title = "{Persistent nonequilibrium effects in generalized Langevin dynamics of nonrelativistic and relativistic particles}",
    eprint = "2301.12450",
    archivePrefix = "arXiv",
    primaryClass = "hep-ph",
    doi = "10.1103/PhysRevE.107.064131",
    journal = "Phys. Rev. E",
    volume = "107",
    number = "6",
    pages = "064131",
    year = "2023"
}

@article{coffey2012langevin,
  title={The {L}angevin equation},
  author={Coffey, W. T. and Kalmykov, Yu. P. and Waldron, J. T.},
  journal={World Scientific Publishing Co., Inc., River Edge, NJ},
  volume={105},
  number={2053912},
  year={2004},
  DOI = {10.1142/5343},
  URL = {https://doi.org/10.1142/5343},
  publisher={World Scientific Series in Contemporary Chemical Physics}
}

@article{kobelev2000fractional,
  title={Fractional Langevin equation to describe anomalous diffusion},
  author={Kobelev, V and Romanov, E},
  journal={Progress of Theoretical Physics Supplement},
  volume={139},
  pages={470--476},
  year={2000},
  publisher={Oxford Academic}
}

@article{li2012spectral,
  title={Spectral approximations to the fractional integral and derivative},
  author={Li, Changpin and Zeng, Fanhai and Liu, Fawang},
  journal={Fractional Calculus and Applied Analysis},
  volume={15},
  number={3},
  pages={383--406},
  year={2012},
  publisher={De Gruyter}
}

@article {mainardi2008fractional,
    AUTHOR = {Mainardi, F. and Pironi, P.},
     TITLE = {The fractional {L}angevin equation: {B}rownian motion
              revisited},
   JOURNAL = {Extracta Math.},
  FJOURNAL = {Extracta Mathematicae},
    VOLUME = {11},
      YEAR = {1996},
    NUMBER = {1},
     PAGES = {140--154},
      ISSN = {0213-8743},
   MRCLASS = {82C40 (33B20 82C31)},
  MRNUMBER = {1424751},
MRREVIEWER = {Carlo Cercignani},
}

@article{MR1611587,
    AUTHOR = {Mainardi, F.},
     TITLE = {Fractional calculus: some basic problems in continuum and
              statistical mechanics},
 SERIES = {CISM Courses and Lect.},
    VOLUME = {378},
     PAGES = {291--348},
 PUBLISHER = {Springer, Vienna},
      YEAR = {1997},
   MRCLASS = {26A33 (33E20 45J05 45K05 73V99)},
  MRNUMBER = {1611587},
MRREVIEWER = {S. L. Kalla},
       DOI = {10.1007/978-3-7091-2664-6\_7},
       URL = {https://doi.org/10.1007/978-3-7091-2664-6_7},
}

@article{MR1611585,
    AUTHOR = {Gorenflo, R. and Mainardi, F.},
     TITLE = {Fractional calculus: integral and differential equations of
              fractional order},
     SERIES = {CISM Courses and Lect.},
    VOLUME = {378},
     PAGES = {223--276},
 PUBLISHER = {Springer, Vienna},
      YEAR = {1997},
   MRCLASS = {26A33 (45E10 45J05)},
  MRNUMBER = {1611585},
MRREVIEWER = {Philip Heywood},
}

@article{li2013finite,
  title={The finite difference methods for fractional ordinary differential equations},
  author={Li, Changpin and Zeng, Fanhai},
  journal={Numerical Functional Analysis and Optimization},
  volume={34},
  number={2},
  pages={149--179},
  year={2013},
  publisher={Taylor \& Francis}
}

@article{miller1993introduction,
  title={An introduction to the fractional calculus and fractional
              differential equations},
  author={Miller, Kenneth S. and Ross, Bertram},
  journal={John Wiley \& Sons, Inc., New York},
  number={1219954},
  pages={xvi+366},
  year={1993},
  publisher={A Wiley-Interscience Publication}
}

@article{bouchaud1990anomalous,
  title={Anomalous diffusion in disordered media: statistical mechanisms, models and physical applications},
  author={Bouchaud, Jean-Philippe and Georges, Antoine},
  journal={Physics reports},
  volume={195},
  number={4-5},
  pages={127--293},
  year={1990},
  publisher={Elsevier}
}

@article{balescu1995anomalous,
  title={Anomalous transport in turbulent plasmas and continuous time random walks},
  author={Balescu, Radu},
  journal={Physical Review E},
  volume={51},
  number={5},
  pages={4807},
  year={1995},
  publisher={APS},
 doi = {10.1103/PhysRevE.51.4807},
}

@article{Oliva:2024rex,
    author = "Oliva, Lucia and Parisi, Gabriele and Greco, Vincenzo and Ruggieri, Marco",
    title = "{Melting of cc{\textasciimacron} and bb{\textasciimacron} pairs in the pre-equilibrium stage of proton-nucleus collisions at the Large Hadron Collider}",
    eprint = "2412.07967",
    archivePrefix = "arXiv",
    primaryClass = "hep-ph",
    doi = "10.1103/nc3z-vns9",
    journal = "Phys. Rev. D",
    volume = "112",
    number = "1",
    pages = "014008",
    year = "2025"
}

@article{metzler2000random,
  title={The random walk's guide to anomalous diffusion: a fractional dynamics approach},
  author={Metzler, Ralf and Klafter, Joseph},
  journal={Physics reports},
  volume={339},
  number={1},
  pages={1--77},
  year={2000},
  publisher={Elsevier}
}

@article{PhysRevLett.85.5655,
  title = {Enhanced Diffusion in Active Intracellular Transport},
  author = {Caspi, Avi and Granek, Rony and Elbaum, Michael},
  journal = {Phys. Rev. Lett.},
  volume = {85},
  issue = {26},
  pages = {5655--5658},
  numpages = {0},
  year = {2000},
  month = {Dec},
  publisher = {American Physical Society},
  doi = {10.1103/PhysRevLett.85.5655},
  url = {https://link.aps.org/doi/10.1103/PhysRevLett.85.5655}
}

@article{PhysRevLett.104.238102,
  title = {Bacterial Chromosomal Loci Move Subdiffusively through a Viscoelastic Cytoplasm},
  author = {Weber, Stephanie C. and Spakowitz, Andrew J. and Theriot, Julie A.},
  journal = {Phys. Rev. Lett.},
  volume = {104},
  issue = {23},
  pages = {238102},
  numpages = {4},
  year = {2010},
  month = {Jun},
  publisher = {American Physical Society},
  doi = {10.1103/PhysRevLett.104.238102},
  url = {https://link.aps.org/doi/10.1103/PhysRevLett.104.238102}
}

@article{sokolov2005diffusion,
  title={From diffusion to anomalous diffusion: a century after Einstein’s Brownian motion},
  author={Sokolov, Igor M and Klafter, Joseph},
  journal={Chaos: An Interdisciplinary Journal of Nonlinear Science},
  volume={15},
  number={2},
  year={2005},
  publisher={AIP Publishing}
}

@article{Prakash:2024irm,
    author = "Prakash, Jai",
    title = "{Subdiffusion of heavy quarks in hot QCD matter by the fractional Langevin equation}",
    eprint = "2406.18714",
    archivePrefix = "arXiv",
    primaryClass = "hep-ph",
    doi = "10.1103/PhysRevC.110.044902",
    journal = "Phys. Rev. C",
    volume = "110",
    number = "4",
    pages = "044902",
    year = "2024"
}

@article{Prakash:2024rdz,
    author = "Prakash, Jai",
    title = "{Anomalous diffusion of the heavy quarks through the fractional Langevin equation}",
    eprint = "2401.03757",
    archivePrefix = "arXiv",
    primaryClass = "hep-ph",
    doi = "10.1103/PhysRevD.109.114004",
    journal = "Phys. Rev. D",
    volume = "109",
    number = "11",
    pages = "114004",
    year = "2024"
}

@article{STAR:2006vcp,
    author = "Adams, J. and others",
    collaboration = "STAR Collaboration",
    title = "{Direct observation of dijets in central Au+Au collisions at s(NN)**(1/2) = 200-GeV}",
    eprint = "nucl-ex/0604018",
    archivePrefix = "arXiv",
    doi = "10.1103/PhysRevLett.97.162301",
    journal = "Phys. Rev. Lett.",
    volume = "97",
    pages = "162301",
    year = "2006"
}

@article{Adams:2005dq,
    author = "Adams, John and others",
    collaboration = "STAR Collaboration",
    title = "{Experimental and theoretical challenges in the search for the quark gluon plasma: The STAR Collaboration's critical assessment of the evidence from RHIC collisions}",
    eprint = "nucl-ex/0501009",
    archivePrefix = "arXiv",
    doi = "10.1016/j.nuclphysa.2005.03.085",
    journal = "Nucl. Phys. A",
    volume = "757",
    pages = "102--183",
    year = "2005"
}

@article{PHENIX:2004vcz,
    author = "Adcox, K. and others",
    collaboration = "PHENIX Collaboration",
    title = "{Formation of dense partonic matter in relativistic nucleus-nucleus collisions at RHIC: Experimental evaluation by the PHENIX collaboration}",
    eprint = "nucl-ex/0410003",
    archivePrefix = "arXiv",
    doi = "10.1016/j.nuclphysa.2005.03.086",
    journal = "Nucl. Phys. A",
    volume = "757",
    pages = "184--283",
    year = "2005"
}

@article{Svetitsky:1987gq,
    author = "Svetitsky, B.",
    title = "{Diffusion of charmed quarks in the quark-gluon plasma}",
    doi = "10.1103/PhysRevD.37.2484",
    journal = "Phys. Rev. D",
    volume = "37",
    pages = "2484--2491",
    year = "1988"
}

@article{BRAHMS:2005gow,
    author = "Arsene, I. and others",
    collaboration = "BRAHMS Collaboration",
    title = "{Centrality dependent particle production at y=0 and y \textasciitilde{} 1 in Au + Au collisions at s(NN)**(1/2) = 200-GeV}",
    eprint = "nucl-ex/0503010",
    archivePrefix = "arXiv",
    doi = "10.1103/PhysRevC.72.014908",
    journal = "Phys. Rev. C",
    volume = "72",
    pages = "014908",
    year = "2005"
}

@article{vanHees:2004gq,
    author = "van Hees, Hendrik and Rapp, Ralf",
    title = "{Thermalization of heavy quarks in the quark-gluon plasma}",
    eprint = "nucl-th/0412015",
    archivePrefix = "arXiv",
    doi = "10.1103/PhysRevC.71.034907",
    journal = "Phys. Rev. C",
    volume = "71",
    pages = "034907",
    year = "2005"
}

@article{Moore:2004tg,
    author = "Moore, Guy D. and Teaney, Derek",
    title = "{How much do heavy quarks thermalize in a heavy ion collision?}",
    eprint = "hep-ph/0412346",
    archivePrefix = "arXiv",
    doi = "10.1103/PhysRevC.71.064904",
    journal = "Phys. Rev. C",
    volume = "71",
    pages = "064904",
    year = "2005"
}

@article{Cao:2011et,
    author = "Cao, Shanshan and Bass, Steffen A.",
    title = "{Thermalization of charm quarks in infinite and finite QGP matter}",
    eprint = "1108.5101",
    archivePrefix = "arXiv",
    primaryClass = "nucl-th",
    doi = "10.1103/PhysRevC.84.064902",
    journal = "Phys. Rev. C",
    volume = "84",
    pages = "064902",
    year = "2011"
}

@article{Das:2024vac,
    author = "Das, Santosh K. and Torres-Rincon, Juan M. and Rapp, Ralf",
    title = "{Charm and Bottom Hadrons in Hot Hadronic Matter}",
    eprint = "2406.13286",
    archivePrefix = "arXiv",
    primaryClass = "hep-ph",
    month = "6",
    year = "2024"
}

@article{Debnath:2023zet,
    author = "Debnath, Manas and Ghosh, Ritesh and Jamal, Mohammad Yousuf and Kurian, Manu and Prakash, Jai",
    title = "{Energy loss of a fast moving parton in Gribov-Zwanziger plasma}",
    eprint = "2311.16005",
    archivePrefix = "arXiv",
    primaryClass = "hep-ph",
    doi = "10.1103/PhysRevD.109.L011503",
    journal = "Phys. Rev. D",
    volume = "109",
    number = "1",
    pages = "L011503",
    year = "2024"
}

@article{Das:2024wht,
    author = "Das, Santosh K. and others",
    title = "{Dynamics of hot QCD matter 2024 {\textemdash} hard probes}",
    eprint = "2412.14026",
    archivePrefix = "arXiv",
    primaryClass = "nucl-ex",
    doi = "10.1142/S0218301325440033",
    journal = "Int. J. Mod. Phys. E",
    volume = "34",
    number = "07",
    pages = "2544003",
    year = "2025"
}

@article{Song:2015sfa,
    author = "Song, Taesoo and Berrehrah, Hamza and Cabrera, Daniel and Torres-Rincon, Juan M. and Tolos, Laura and Cassing, Wolfgang and Bratkovskaya, Elena",
    title = "{Tomography of the Quark-Gluon-Plasma by Charm Quarks}",
    eprint = "1503.03039",
    archivePrefix = "arXiv",
    primaryClass = "nucl-th",
    doi = "10.1103/PhysRevC.92.014910",
    journal = "Phys. Rev. C",
    volume = "92",
    number = "1",
    pages = "014910",
    year = "2015"
}

@article{Andronic:2015wma,
    author = "Andronic, A. and others",
    title = "{Heavy-flavour and quarkonium production in the LHC era: from proton\textendash{}proton to heavy-ion collisions}",
    eprint = "1506.03981",
    archivePrefix = "arXiv",
    primaryClass = "nucl-ex",
    doi = "10.1140/epjc/s10052-015-3819-5",
    journal = "Eur. Phys. J. C",
    volume = "76",
    number = "3",
    pages = "107",
    year = "2016"
}

@article{Dong:2019unq,
    author = "Dong, Xin and Greco, Vincenzo",
    title = "{Heavy quark production and properties of Quark\textendash{}Gluon Plasma}",
    doi = "10.1016/j.ppnp.2018.08.001",
    journal = "Prog. Part. Nucl. Phys.",
    volume = "104",
    pages = "97--141",
    year = "2019"
}

@article{ALICE:2010khr,
    author = "Aamodt, K and others",
    collaboration = "ALICE Collaboration",
    title = "{Charged-particle multiplicity density at mid-rapidity in central Pb-Pb collisions at $\sqrt{s_{NN}} = 2.76$ TeV}",
    eprint = "1011.3916",
    archivePrefix = "arXiv",
    primaryClass = "nucl-ex",
    reportNumber = "CERN-PH-EP-2010-060",
    doi = "10.1103/PhysRevLett.105.252301",
    journal = "Phys. Rev. Lett.",
    volume = "105",
    pages = "252301",
    year = "2010"
}

@article{Aarts:2016hap,
    author = "Aarts, G. and others",
    title = "{Heavy-flavor production and medium properties in high-energy nuclear collisions - What next?}",
    eprint = "1612.08032",
    archivePrefix = "arXiv",
    primaryClass = "nucl-th",
    doi = "10.1140/epja/i2017-12282-9",
    journal = "Eur. Phys. J. A",
    volume = "53",
    number = "5",
    pages = "93",
    year = "2017"
}

@article{Plumari:2017ntm,
    author = "Plumari, Salvatore and Minissale, Vincenzo and Das, Santosh K. and Coci, G. and Greco, V.",
    title = "{Charmed Hadrons from Coalescence plus Fragmentation in relativistic nucleus-nucleus collisions at RHIC and LHC}",
    eprint = "1712.00730",
    archivePrefix = "arXiv",
    primaryClass = "hep-ph",
    doi = "10.1140/epjc/s10052-018-5828-7",
    journal = "Eur. Phys. J. C",
    volume = "78",
    number = "4",
    pages = "348",
    year = "2018"
}

@article{Greco:2015uhf,
    author = "Greco, Vincenzo and Scardina, Francesco and Das, Santosh K and Plumari, Salvatore",
    title = "{Heavy Quark Dynamics in QGP: Boltzmann vs Langevin}",
    doi = "10.16943/ptinsa/2015/v81i1/48068",
    journal = "Proc. Indian Natl. Sci. Acad.",
    volume = "81",
    number = "1",
    pages = "186--198",
    year = "2015"
}

@article{vanHees:2007me,
    author = "van Hees, H. and Mannarelli, M. and Greco, V. and Rapp, R.",
    title = "{Nonperturbative heavy-quark diffusion in the quark-gluon plasma}",
    eprint = "0709.2884",
    archivePrefix = "arXiv",
    primaryClass = "hep-ph",
    doi = "10.1103/PhysRevLett.100.192301",
    journal = "Phys. Rev. Lett.",
    volume = "100",
    pages = "192301",
    year = "2008"
}

@article{Das:2015ana,
    author = "Das, Santosh K. and Scardina, Francesco and Plumari, Salvatore and Greco, Vincenzo",
    title = "{Toward a solution to the $R_{AA}$ and $v_2$ puzzle for heavy quarks}",
    eprint = "1502.03757",
    archivePrefix = "arXiv",
    primaryClass = "nucl-th",
    doi = "10.1016/j.physletb.2015.06.003",
    journal = "Phys. Lett. B",
    volume = "747",
    pages = "260--264",
    year = "2015"
}

@incollection{Rapp:2009my,
    author = "Rapp, Ralf and van Hees, Hendrik",
    title = "{Heavy Quarks in the Quark-Gluon Plasma}",
    booktitle = "{Quark-Gluon Plasma 4}",
    editor = "Hwa, Rudolph C. and Wang, Xin-Nian",
    publisher = "World Scientific",
    location = "Singapore",
    eprint = "0903.1096",
    archivePrefix = "arXiv",
    primaryClass = "hep-ph",
    doi = "10.1142/9789814293297_0003",
    pages = "111--206",
    year = "2010"
}

@article{Cao:2013ita,
    author = "Cao, Shanshan and Qin, Guang-You and Bass, Steffen A.",
    title = "{Heavy-quark dynamics and hadronization in ultrarelativistic heavy-ion collisions: Collisional versus radiative energy loss}",
    eprint = "1308.0617",
    archivePrefix = "arXiv",
    primaryClass = "nucl-th",
    doi = "10.1103/PhysRevC.88.044907",
    journal = "Phys. Rev. C",
    volume = "88",
    pages = "044907",
    year = "2013"
}

@article{Das:2013kea,
    author = "Das, Santosh K. and Scardina, Francesco and Plumari, Salvatore and Greco, Vincenzo",
    title = "{Heavy-flavor in-medium momentum evolution: Langevin versus Boltzmann approach}",
    eprint = "1312.6857",
    archivePrefix = "arXiv",
    primaryClass = "nucl-th",
    doi = "10.1103/PhysRevC.90.044901",
    journal = "Phys. Rev. C",
    volume = "90",
    pages = "044901",
    year = "2014"
}

@article{scardina2017estimating,
  title={Estimating the charm quark diffusion coefficient and thermalization time from D meson spectra at energies available at the BNL Relativistic Heavy Ion Collider and the CERN Large Hadron Collider},
  author={Scardina, Francesco and Das, Santosh K and Minissale, Vincenzo and Plumari, Salvatore and Greco, Vincenzo},
  journal={Physical Review C},
  volume={96},
  number={4},
  pages={044905},
  year={2017},
  publisher={APS}
}

@article{Akamatsu:2008ge,
    author = "Akamatsu, Yukinao and Hatsuda, Tetsuo and Hirano, Tetsufumi",
    title = "{Heavy Quark Diffusion with Relativistic Langevin Dynamics in the Quark-Gluon Fluid}",
    eprint = "0809.1499",
    archivePrefix = "arXiv",
    primaryClass = "hep-ph",
    reportNumber = "TKYNT-08-14",
    doi = "10.1103/PhysRevC.79.054907",
    journal = "Phys. Rev. C",
    volume = "79",
    pages = "054907",
    year = "2009"
}

@article{ZACCONE2024116483,
title = {Theory of heavy-quarks contribution to the quark-gluon plasma viscosity},
journal = {Nuclear Physics B},
volume = {1000},
pages = {116483},
year = {2024},
issn = {0550-3213},
doi = {10.1016/j.nuclphysb.2024.116483},
url = {https://www.sciencedirect.com/science/article/pii/S055032132400049X},
author = {Alessio Zaccone},

}

@article{Das:2025yxy,
    author = "Das, Santosh K. and Soloveva, Olga and Song, Taesoo and Bratkovskaya, Elena",
    title = "{Influence of electromagnetic fields on the generation of the directed and elliptic flows of heavy quarks in relativistic heavy-ion collisions}",
    eprint = "2507.22620",
    archivePrefix = "arXiv",
    primaryClass = "nucl-th",
    doi = "10.1103/yppc-ts4n",
    journal = "Phys. Rev. C",
    volume = "112",
    number = "6",
    pages = "064901",
    year = "2025"
}

@article{Minissale:2023dct,
    author = "Minissale, Vincenzo and Plumari, Salvatore and Sun, Yifeng and Greco, Vincenzo",
    title = "{Multi-charmed and singled charmed hadrons from coalescence: yields and ratios in different collision systems at LHC}",
    eprint = "2305.03687",
    archivePrefix = "arXiv",
    primaryClass = "hep-ph",
    doi = "10.1140/epjc/s10052-024-12571-6",
    journal = "Eur. Phys. J. C",
    volume = "84",
    number = "3",
    pages = "228",
    year = "2024"
}

@article{Sambataro:2025pop,
    author = "Sambataro, Maria Lucia and Plumari, Salvatore and Das, Santosh K. and Greco, Vincenzo",
    title = "{Probing the QGP through $p_T$-differential radial flow of heavy quarks}",
    eprint = "2510.19448",
    archivePrefix = "arXiv",
    primaryClass = "hep-ph",
    doi = "10.1103/v4mq-n7vb",
    journal = "Phys. Rev. Lett.",
    volume = "136",
    pages = "212302",
    year = "2026"
}

@article{PhysRevLett.130.231902,
  title = {Heavy Quark Diffusion from $2+1$ Flavor Lattice QCD with 320 MeV Pion Mass},
  author = {Altenkort, Luis and Kaczmarek, Olaf and Larsen, Rasmus and Mukherjee, Swagato and Petreczky, Peter and Shu, Hai-Tao and Stendebach, Simon},
  collaboration = {HotQCD Collaboration},
  journal = {Phys. Rev. Lett.},
  volume = {130},
  issue = {23},
  pages = {231902},
  numpages = {6},
  year = {2023},
  month = {Jun},
  publisher = {American Physical Society},
  doi = {10.1103/PhysRevLett.130.231902},
  url = {https://link.aps.org/doi/10.1103/PhysRevLett.130.231902}
}

@article{Cao:2016gvr,
    author = "Cao, Shanshan and Luo, Tan and Qin, Guang-You and Wang, Xin-Nian",
    title = "{Linearized Boltzmann transport model for jet propagation in the quark-gluon plasma: Heavy quark evolution}",
    eprint = "1605.06447",
    archivePrefix = "arXiv",
    primaryClass = "nucl-th",
    doi = "10.1103/PhysRevC.94.014909",
    journal = "Phys. Rev. C",
    volume = "94",
    number = "1",
    pages = "014909",
    year = "2016"
}

@article{Gossiaux:2008jv,
    author = "Gossiaux, P. B. and Aichelin, J.",
    title = "{Towards an understanding of the RHIC single electron data}",
    eprint = "0802.2525",
    archivePrefix = "arXiv",
    primaryClass = "hep-ph",
    doi = "10.1103/PhysRevC.78.014904",
    journal = "Phys. Rev. C",
    volume = "78",
    pages = "014904",
    year = "2008"
}

@article{Scardina:2017ipo,
    author = "Scardina, Francesco and Das, Santosh K. and Minissale, Vincenzo and Plumari, Salvatore and Greco, Vincenzo",
    title = "{Estimating the charm quark diffusion coefficient and thermalization time from D meson spectra at energies available at the BNL Relativistic Heavy Ion Collider and the CERN Large Hadron Collider}",
    eprint = "1707.05452",
    archivePrefix = "arXiv",
    primaryClass = "nucl-th",
    doi = "10.1103/PhysRevC.96.044905",
    journal = "Phys. Rev. C",
    volume = "96",
    number = "4",
    pages = "044905",
    year = "2017"
}

@article{cacciari2005qcd,
    author = "Cacciari, Matteo and Nason, Paolo and Vogt, Ramona",
    title = "{QCD predictions for charm and bottom production at RHIC}",
    eprint = "hep-ph/0502203",
    archivePrefix = "arXiv",
    reportNumber = "LPTHE-05-03, BIOCOCCA-FT-05-4, LBNL-57063, BICOCCA-FT-05-4",
    doi = "10.1103/PhysRevLett.95.122001",
    journal = "Phys. Rev. Lett.",
    volume = "95",
    pages = "122001",
    year = "2005"
}

@article{Cao:2015hia,
    author = "Cao, Shanshan and Qin, Guang-You and Bass, Steffen A.",
    title = "{Energy loss, hadronization and hadronic interactions of heavy flavors in relativistic heavy-ion collisions}",
    eprint = "1505.01413",
    archivePrefix = "arXiv",
    primaryClass = "nucl-th",
    doi = "10.1103/PhysRevC.92.024907",
    journal = "Phys. Rev. C",
    volume = "92",
    number = "2",
    pages = "024907",
    year = "2015"
}

@article{Cacciari:2012ny,
    author = "Cacciari, Matteo and Frixione, Stefano and Houdeau, Nicolas and Mangano, Michelangelo L. and Nason, Paolo and Ridolfi, Giovanni",
    title = "{Theoretical predictions for charm and bottom production at the LHC}",
    eprint = "1205.6344",
    archivePrefix = "arXiv",
    primaryClass = "hep-ph",
    reportNumber = "CERN-PH-TH-2011-227",
    doi = "10.1007/JHEP10(2012)137",
    journal = "JHEP",
    volume = "10",
    pages = "137",
    year = "2012"

}

@article{vitali2018langevin,
  title={Langevin equation in complex media and anomalous diffusion},
  author={Vitali, Silvia and Sposini, Vittoria and Sliusarenko, Oleksii and Paradisi, Paolo and Castellani, Gastone and Pagnini, Gianni},
  journal={Journal of The Royal Society Interface},
  volume={15},
  number={145},
  pages={20180282},
  year={2018},
  publisher={The Royal Society}
}

@article{oliveira2019anomalous,
  title={Anomalous diffusion: A basic mechanism for the evolution of inhomogeneous systems},
  author={Oliveira, Fernando A and Ferreira, Rogelma MS and Lapas, Luciano C and Vainstein, Mendeli H},
  journal={Frontiers in Physics},
  volume={7},
  pages={18},
  year={2019},
  publisher={Frontiers Media SA}
}

@article{guo2013numerics,
  title={Numerics for the fractional Langevin equation driven by the fractional Brownian motion},
  author={Guo, Peng and Zeng, Caibin and Li, Changpin and Chen, YangQuan},
  journal={Fractional Calculus and Applied Analysis},
  volume={16},
  number={1},
  pages={123--141},
  year={2013},
  publisher={Versita}
}

@article{10.1111/j.1365-246X.1967.tb02303.x, author = {Caputo, Michele}, title = "{Linear Models of Dissipation whose Q is almost Frequency Independent—II}", journal = {Geophysical Journal International}, volume = {13}, number = {5}, pages = {529-539}, year = {1967}, month = {11}, issn = {0956-540X}, doi = {10.1111/j.1365-246X.1967.tb02303.x},  }

@article{carpinteri2014fractals,
  author = "Carpinteri, Alberto and Mainardi, Francesco",
  title = "{Fractals and Fractional Calculus in Continuum Mechanics}",
  journal = "Fractals and Fractional Calculus in Continuum Mechanics",  
  volume = "378",
  year = "2014",
}

@article{Agrawal_2007,
doi = {10.1088/1751-8113/40/24/003},
url = {https://dx.doi.org/10.1088/1751-8113/40/24/003},
year = {2007},
month = {may},
publisher = {},
volume = {40},
number = {24},
pages = {6287},
author = {O P Agrawal},
title = {Fractional variational calculus in terms of Riesz fractional derivatives},
journal = {Journal of Physics A: Mathematical and Theoretical},
}

@article{Feller1971,
  title = {An Introduction to Probability Theory and Its Application Vol II},
  author = {Feller, W.},
  journal = {John Wiley and Sons},
  year = {1971},
  url={https://books.google.co.in/books?id=uPKXwAEACAAJ},
  
}

@article{PhysRevE.107.024105,
  title = {Langevin picture of anomalous diffusion processes in expanding medium},
  author = {Wang, Xudong and Chen, Yao},
  journal = {Phys. Rev. E},
  volume = {107},
  issue = {2},
  pages = {024105},
  numpages = {14},
  year = {2023},
  month = {Feb},
  publisher = {American Physical Society},
  doi = {10.1103/PhysRevE.107.024105},
  url = {https://link.aps.org/doi/10.1103/PhysRevE.107.024105}
}

@article{lutz2001fractional,
  title = {Fractional Langevin equation},
  author = {Lutz, Eric},
  journal = {Phys. Rev. E},
  volume = {64},
  issue = {5},
  pages = {051106},
  numpages = {4},
  year = {2001},
  month = {Oct},
  publisher = {American Physical Society},
  doi = {10.1103/PhysRevE.64.051106},
  url = {https://link.aps.org/doi/10.1103/PhysRevE.64.051106}
}

@article{PhysRevC.86.014903,
  title = {Heavy-quark diffusion and hadronization in quark-gluon plasma},
  author = {He, Min and Fries, Rainer J. and Rapp, Ralf},
  journal = {Phys. Rev. C},
  volume = {86},
  issue = {1},
  pages = {014903},
  numpages = {15},
  year = {2012},
  month = {Jul},
  publisher = {American Physical Society},
  doi = {10.1103/PhysRevC.86.014903},
  url = {https://link.aps.org/doi/10.1103/PhysRevC.86.014903}
}

@article{PhysRevC.86.034905,
  title = {MARTINI event generator for heavy quarks: Initialization, parton evolution, and hadronization},
  author = {Young, Clint and Schenke, Bj\"orn and Jeon, Sangyong and Gale, Charles},
  journal = {Phys. Rev. C},
  volume = {86},
  pages = {034905},
  numpages = {7},
  year = {2012},
  publisher = {American Physical Society},
  doi = {10.1103/PhysRevC.86.034905},
  url = {https://link.aps.org/doi/10.1103/PhysRevC.86.034905}
}

@article{He:2022ywp,
    author = "He, Min and van Hees, Hendrik and Rapp, Ralf",
    title = "{Heavy-quark diffusion in the quark\textendash{}gluon plasma}",
    eprint = "2204.09299",
    archivePrefix = "arXiv",
    primaryClass = "hep-ph",
    doi = "10.1016/j.ppnp.2023.104020",
    journal = "Prog. Part. Nucl. Phys.",
    volume = "130",
    pages = "104020",
    year = "2023"
}

@article{Bhattacharyya:2024hku,
    author = "Bhattacharyya, Trambak and Megias, Eugenio and Deppman, Airton",
    title = "{Jet quenching of the heavy quarks in the quark-gluon plasma and the nonadditive statistics}",
    eprint = "2405.15679",
    archivePrefix = "arXiv",
    primaryClass = "hep-ph",
    doi = "10.1016/j.physletb.2024.138907",
    journal = "Phys. Lett. B",
    volume = "856",
    pages = "138907",
    year = "2024"
}

@article{PhysRevC.84.044901,
  title = {Momentum dependence of drag coefficients and heavy flavor suppression in quark gluon plasma},
  author = {Mazumder, Surasree and Bhattacharyya, Trambak and Alam, Jan-e and Das, Santosh K.},
  journal = {Phys. Rev. C},
  volume = {84},
  issue = {4},
  pages = {044901},
  numpages = {6},
  year = {2011},
  month = {Oct},
  publisher = {American Physical Society},
  doi = {10.1103/PhysRevC.84.044901},
  url = {https://link.aps.org/doi/10.1103/PhysRevC.84.044901}
}

@article{Zhang:2022fum,
    author = "Zhang, Chao and Zheng, Liang and Shi, Shusu and Lin, Zi-Wei",
    title = "{Resolving the RpA and v2 puzzle of D0 mesons in p{\ensuremath{-}}Pb collisions at the LHC}",
    eprint = "2210.07767",
    archivePrefix = "arXiv",
    primaryClass = "nucl-th",
    doi = "10.1016/j.physletb.2023.138219",
    journal = "Phys. Lett. B",
    volume = "846",
    pages = "138219",
    year = "2023"
}

@article{Sun:2023adv,
    author = "Sun, Yifeng and Plumari, Salvatore and Das, Santosh K.",
    title = "{Exploring the effects of electromagnetic fields and tilted bulk distribution on directed flow of D mesons in small systems}",
    eprint = "2304.12792",
    archivePrefix = "arXiv",
    primaryClass = "nucl-th",
    doi = "10.1016/j.physletb.2023.138043",
    journal = "Phys. Lett. B",
    volume = "843",
    pages = "138043",
    year = "2023"
}

@article{Plumari:2019hzp,
    author = "Plumari, Salvatore and Coci, Gabriele and Minissale, Vincenzo and Das, Santosh K. and Sun, Yifeng and Greco, Vincenzo",
    title = "{Heavy - light flavor correlations of anisotropic flows at LHC energies within event-by-event transport approach}",
    eprint = "1912.09350",
    archivePrefix = "arXiv",
    primaryClass = "hep-ph",
    doi = "10.1016/j.physletb.2020.135460",
    journal = "Phys. Lett. B",
    volume = "805",
    pages = "135460",
    year = "2020"
}

@article{du2023accelerated,
  title={Accelerated quantum circuit Monte-Carlo simulation for heavy quark thermalization},
  author={Du, Xiaojian and Qian, Wenyang},
  journal={Phys. Rev. D},
  volume={109},
  pages={076025},
  year={2024},
  doi={10.1103/PhysRevD.109.076025},
  eprint={2312.16294},
  archivePrefix={arXiv},
  primaryClass={hep-ph}
}

@article{MR361633,
    author = "Oldham, Keith B. and Spanier, Jerome",
    title = "{The fractional calculus}",
    eprint = "2105.14296",
    journal = "Mathematics in Science and Engineering",
    volume = "111",
    number = "361633",
    year = "1974"
}

@article{PhysRevE.108.034113,
  title = {Anomalous diffusion, nonergodicity, non-Gaussianity, and aging of fractional Brownian motion with nonlinear clocks},
  author = {Liang, Yingjie and Wang, Wei and Metzler, Ralf and Cherstvy, Andrey G.},
  journal = {Phys. Rev. E},
  volume = {108},
  issue = {3},
  pages = {034113},
  numpages = {24},
  year = {2023},
  month = {Sep},
  publisher = {American Physical Society},
  doi = {10.1103/PhysRevE.108.034113},
  url = {https://link.aps.org/doi/10.1103/PhysRevE.108.034113}
}

@article{PhysRevLett.125.240606,
  title = {Fractional Advection-Diffusion-Asymmetry Equation},
  author = {Wang, Wanli and Barkai, Eli},
  journal = {Phys. Rev. Lett.},
  volume = {125},
  issue = {24},
  pages = {240606},
  numpages = {6},
  year = {2020},
  month = {Dec},
  publisher = {American Physical Society},
  doi = {10.1103/PhysRevLett.125.240606},
  url = {https://link.aps.org/doi/10.1103/PhysRevLett.125.240606}
}

@article{PhysRevE.104.024115,
  title = {Inertia triggers nonergodicity of fractional Brownian motion},
  author = {Cherstvy, Andrey G. and Wang, Wei and Metzler, Ralf and Sokolov, Igor M.},
  journal = {Phys. Rev. E},
  volume = {104},
  issue = {2},
  pages = {024115},
  numpages = {12},
  year = {2021},
  month = {Aug},
  publisher = {American Physical Society},
  doi = {10.1103/PhysRevE.104.024115},
  url = {https://link.aps.org/doi/10.1103/PhysRevE.104.024115}
}

@article{Altenkort:2023oms,
    author = "Altenkort, Luis and Kaczmarek, Olaf and Larsen, Rasmus and Mukherjee, Swagato and Petreczky, Peter and Shu, Hai-Tao and Stendebach, Simon",
    collaboration = "HotQCD Collaboration",
    title = "{Heavy Quark Diffusion from 2+1 Flavor Lattice QCD with 320 MeV Pion Mass}",
    eprint = "2302.08501",
    archivePrefix = "arXiv",
    primaryClass = "hep-lat",
    doi = "10.1103/PhysRevLett.130.231902",
    journal = "Phys. Rev. Lett.",
    volume = "130",
    number = "23",
    pages = "231902",
    year = "2023"
}

@article{Altenkort:2023eav,
    author = "Altenkort, Luis and de la Cruz, David and Kaczmarek, Olaf and Larsen, Rasmus and Moore, Guy D. and Mukherjee, Swagato and Petreczky, Peter and Shu, Hai-Tao and Stendebach, Simon",
    collaboration = "HotQCD Collaboration",
    title = "{Quark Mass Dependence of Heavy Quark Diffusion Coefficient from Lattice QCD}",
    eprint = "2311.01525",
    archivePrefix = "arXiv",
    primaryClass = "hep-lat",
    doi = "10.1103/PhysRevLett.132.051902",
    journal = "Phys. Rev. Lett.",
    volume = "132",
    number = "5",
    pages = "051902",
    year = "2024"
}

@article{Xu:2017obm,
    author = "Xu, Yingru and Bernhard, Jonah E. and Bass, Steffen A. and Nahrgang, Marlene and Cao, Shanshan",
    title = "{Data-driven analysis for the temperature and momentum dependence of the heavy-quark diffusion coefficient in relativistic heavy-ion collisions}",
    eprint = "1710.00807",
    archivePrefix = "arXiv",
    primaryClass = "nucl-th",
    doi = "10.1103/PhysRevC.97.014907",
    journal = "Phys. Rev. C",
    volume = "97",
    number = "1",
    pages = "014907",
    year = "2018"
}

@ARTICLE{2017JSP...169..316L,
       author = {{Li}, Lei and {Liu}, Jian-Guo and {Lu}, Jianfeng},
        title = "{Fractional Stochastic Differential Equations Satisfying Fluctuation-Dissipation Theorem}",
      journal = {Journal of Statistical Physics},
         year = 2017,
        month = oct,
       volume = {169},
       number = {2},
        pages = {316-339},
          doi = {10.1007/s10955-017-1866-z},
archivePrefix = {arXiv},
       eprint = {1612.04274},
 primaryClass = {math.PR},
       adsurl = {https://ui.adsabs.harvard.edu/abs/2017JSP...169..316L}
}

@article{Wiener1930,
  author  = {Wiener, Norbert},
  title   = {Generalized Harmonic Analysis},
  journal = {Acta Mathematica},
  volume  = {55},
  pages   = {117--258},
  year    = {1930},
  doi     = {10.1007/BF02546511}
}

@article{Khinchin1934,
  author  = {Khintchine, Aleksandr},
  title   = {Korrelationstheorie der station{\"a}ren stochastischen Prozesse},
  journal = {Mathematische Annalen},
  volume  = {109},
  pages   = {604--615},
  year    = {1934},
  doi     = {10.1007/BF01449156}
}

@book{PapoulisPillai2002,
  author    = {Papoulis, Athanasios and Pillai, S. Unnikrishna},
  title     = {Probability, Random Variables and Stochastic Processes},
  edition   = {4th},
  publisher = {McGraw-Hill},
  address   = {New York},
  year      = {2002}
}

@book{Gardiner,
  author    = {Gardiner, Crispin W.},
  title     = {Handbook of Stochastic Methods for Physics, Chemistry and the Natural Sciences},
  series    = {Springer Series in Synergetics},
  publisher = {Springer-Verlag},
  address   = {Berlin},
}

@article{MandelbrotVanNess1968,
  author  = {Mandelbrot, Benoit B. and Van Ness, John W.},
  title   = {Fractional {B}rownian Motions, Fractional Noises and Applications},
  journal = {SIAM Review},
  volume  = {10},
  pages   = {422--437},
  year    = {1968},
  doi     = {10.1137/1010093}
}

@book{Yaglom1987,
  author    = {Yaglom, A. M.},
  title     = {Correlation Theory of Stationary and Related Random Functions I: Basic Results},
  publisher = {Springer-Verlag},
  address   = {New York},
  year      = {1987}
}

@article{Nair:2026tqy,
    author = "Nair, Monisha and Torres-Rincon, Juan and Das, Santosh K.",
    title = "{Non-Markovian heavy-quark equilibration and equilibrium correlation function in a thermal medium}",
    eprint = "2607.22403",
    archivePrefix = "arXiv",
    primaryClass = "hep-ph",
    doi = "10.48550/arXiv.2607.22403",
    journal = "arXiv:2607.22403",
    year = "2026"
}

\end{document}


\maketitle

\section{Fractional fluctuation-dissipation theorem}

\ \ \ \ Taking the Fourier transform of the fractional Langevin equation,
\begin{align}
\tau_{p}^{\,\alpha-1}(i\omega)^{\alpha}\,\widetilde{p}(\omega)
  = -\gamma\,\widetilde{p}(\omega) + \widetilde{\xi}(\omega).
\label{eq:FLE_Fourier}
\end{align}
The Fourier transformation of the heavy quark momentum becomes,
\begin{align}
\widetilde{p}(\omega)
  = \frac{\widetilde{\xi}(\omega)}
         {\gamma+\tau_{p}^{\,\alpha-1}(i\omega)^{\alpha}}
  \equiv H(\omega)\,\widetilde{\xi}(\omega),
\qquad
H(\omega)=\frac{1}{\gamma+\tau_{p}^{\,\alpha-1}(i\omega)^{\alpha}}.
\label{eq:p_Fourier}
\end{align}
The momentum power spectral density ($S_{p}(\omega)$) is the distribution of momentum fluctuations of the heavy quark over frequency. For a stationary momentum process, the Fourier transform is taken over a finite interval of duration $\mathcal{T}$, the mean-squared modulus is divided by $\mathcal{T}$, and the long-time limit is taken\cite{Wiener1930,Khinchin1934,PapoulisPillai2002},
\begin{align}
S_{p}(\omega)
  = \lim_{\mathcal{T}\to\infty}\frac{1}{\mathcal{T}}
\left\langle\left|\widetilde{p}(\omega)\right|^{2}\right\rangle,
\label{eq:Sp_def}
\end{align}
where $\widetilde{p}(\omega)=\int_{-\mathcal{T}/2}^{\mathcal{T}/2} p(t)\,e^{-i\omega t}\,dt$ . Expanding the squared modulus as a double time integral and using stationarity, $\langle p(t)\,p(t')\rangle = C_{p}(t-t')$ with the momentum autocorrelation $C_{p}(\tau)=\langle p(t)\,p(t+\tau)\rangle$, the integrand depends only on the lag,
\begin{align}
\left\langle\left|\widetilde{p}(\omega)\right|^{2}\right\rangle
  = \int_{-\mathcal{T}/2}^{\mathcal{T}/2}\!\!\int_{-\mathcal{T}/2}^{\mathcal{T}/2}
    C_{p}(t-t')\,e^{-i\omega(t-t')}\,dt\,dt'.
\label{eq:windowed_avg}
\end{align}
Introducing the lag $\tau=t-t'$ and performing the $t'$ integration, the double integral reduces to
\begin{align}
\frac{1}{\mathcal{T}}\left\langle\left|\widetilde{p}(\omega)\right|^{2}\right\rangle
  = \int_{-\mathcal{T}}^{\mathcal{T}}
    \!\Big(1-\frac{|\tau|}{\mathcal{T}}\Big)\,C_{p}(\tau)\,e^{-i\omega\tau}\,d\tau.
\label{eq:bartlett}
\end{align}
For a correlation function that decays on a finite timescale, the factor $(1-|\tau|/\mathcal{T})\to1$ for every finite $\tau$ as $\mathcal{T}\to\infty$, and the integration limits extend to $\pm\infty$, yielding
\begin{align}
S_{p}(\omega) = \int_{-\infty}^{\infty} C_{p}(\tau)\,e^{-i\omega\tau}\,d\tau.
\label{eq:WK_theorem}
\end{align}
The same construction applied to the stochastic force defines the noise power spectral density $S_{\xi}(\omega)=\lim_{\mathcal{T}\to\infty}\tfrac{1}{\mathcal{T}}\langle|\widetilde{\xi}(\omega)|^{2}\rangle$. Substituting the  relation in Eq.~\eqref{eq:p_Fourier} into Eq.~\eqref{eq:Sp_def} and taking the deterministic factor $|H(\omega)|^{2}$ outside the average yields \cite{PapoulisPillai2002,Gardiner}
\begin{align}
S_{p}(\omega) = |H(\omega)|^{2}\,S_{\xi}(\omega).
\label{eq:Sp_Sxi}
\end{align}
The thermal noise is taken to be additive and white, so that its spectral power density is frequency independent, $S_{\xi}(\omega) = \int_{-\infty}^{\infty}\langle\xi(t)\,\xi(t+\tau)\rangle\,e^{-i\omega\tau}\,d\tau = 2\mathcal{D}$, and the $S_{p}(\omega)$ is
\begin{align}
S_{p}(\omega)
  = \frac{2\mathcal{D}}
         {\bigl|\gamma+\tau_{p}^{\,\alpha-1}(i\omega)^{\alpha}\bigr|^{2}}.
\label{eq:Sp_white}
\end{align}
With the principal branch $(i\omega)^{\alpha}=|\omega|^{\alpha}\bigl[\cos\tfrac{\pi\alpha}{2}+i\,\mathrm{sgn}(\omega)\sin\tfrac{\pi\alpha}{2}\bigr]$, the squared modulus of the denominator follows from $|a+ib|^{2}=a^{2}+b^{2}$,
\begin{align}
\bigl|\gamma+\tau_{p}^{\,\alpha-1}(i\omega)^{\alpha}\bigr|^{2}
  = \gamma^{2}
   +2\gamma\,\tau_{p}^{\,\alpha-1}|\omega|^{\alpha}\cos\tfrac{\pi\alpha}{2}
   +\tau_{p}^{\,2(\alpha-1)}|\omega|^{2\alpha}.
\label{eq:denom_expand}
\end{align}
The stationary momentum variance, reached in the long-time limit, is the zero-lag value of the autocorrelation, $\lim_{t\to\infty}\langle p^{2}(t)\rangle=C_{p}(0)$. Inverting Eq.~\eqref{eq:WK_theorem} and evaluating the result at $\tau=0$ gives the equilibrium variance as the frequency integral of the power spectral density,
\begin{align}
\lim_{t\to\infty}\langle p^{2}(t)\rangle
  = \frac{1}{2\pi}\int_{-\infty}^{\infty} S_{p}(\omega)\,d\omega.
\label{eq:p2_twosided}
\end{align}
The integral over the full axis equals twice that over the positive axis; factoring $\gamma^{2}$ from the denominator,
\begin{align}
\lim_{t\to\infty}\langle p^{2}(t)\rangle
  = \frac{2\mathcal{D}}{\pi\gamma^{2}}
     \!\int_{0}^{\infty}\!
     \frac{d\omega}
          {1+2\bigl(\tfrac{\tau_{p}^{\alpha-1}}{\gamma}\bigr)\omega^{\alpha}\cos\frac{\pi\alpha}{2}
           +\bigl(\tfrac{\tau_{p}^{\alpha-1}}{\gamma}\bigr)^{2}\omega^{2\alpha}}.
\label{eq:WK_integral}
\end{align}
The substitution $u=(\tau_{p}^{\alpha-1}/\gamma)\,\omega^{\alpha}$, with $d\omega=\tfrac{1}{\alpha}(\gamma/\tau_{p}^{\alpha-1})^{1/\alpha}\,u^{1/\alpha-1}\,du$, casts Eq.~\eqref{eq:WK_integral} into the form
\begin{align}
\lim_{t\to\infty}\langle p^{2}(t)\rangle
  = \frac{2\mathcal{D}}{\pi\alpha}\,\gamma^{1/\alpha-2}\,\tau_{p}^{-(\alpha-1)/\alpha}
    \int_{0}^{\infty}\frac{u^{1/\alpha-1}}{1+2u\cos\frac{\pi\alpha}{2}+u^{2}}\,du,
\label{eq:p2_substituted}
\end{align}
with $\displaystyle\int_{0}^{\infty}\frac{u^{m-1}}{1+2u\cos\theta+u^{2}}\,du=\frac{\pi\,\sin[(1-m)\theta]}{\sin\theta\,\sin(\pi m)}$ for $0<m<2$. Setting $m=1/\alpha$ and $\theta=\pi\alpha/2$ gives $(1-m)\theta=\pi(\alpha-1)/2$, and the variance reduces to
\begin{align}
\lim_{t\to\infty}\langle p^{2}(t)\rangle
  = \frac{2\mathcal{D}}{\alpha}\,\gamma^{1/\alpha-2}\,\tau_{p}^{-(\alpha-1)/\alpha}\,
    \frac{\sin\frac{\pi(\alpha-1)}{2}}{\sin\frac{\pi\alpha}{2}\,\sin\frac{\pi}{\alpha}}.
\label{eq:p2_closed}
\end{align}
Imposing the equipartition condition $\lim_{t\to\infty}\langle p^{2}(t)\rangle=MT$ for the case of 1 dimension and solving for $\mathcal{D}$ yields the fractional fluctuation--dissipation relation
\begin{align}
\mathcal{D}(\alpha,\tau_{p})
  = K_{\alpha}\,MT\,\gamma^{\,2-1/\alpha}\,\tau_{p}^{(\alpha-1)/\alpha},
\qquad
K_{\alpha}=\frac{\alpha}{2}\,
  \frac{\sin\frac{\pi\alpha}{2}\,\sin\frac{\pi}{\alpha}}{\sin\frac{\pi(\alpha-1)}{2}}.
\label{eq:FFDT}
\end{align}
In the limit $\alpha\to1$, $K_{\alpha}\to1$ and the standard fluctuation--dissipation relation $\mathcal{D}=\gamma MT$ is recovered.

\section{Mean-square displacement: integral representation}

 \ \ \ \ For a non-relativistic heavy  quark of mass $M$, the position is related to the momentum through the relation,
\begin{align}
\tau_{p}^{\,\beta-1}\,{}^{C}\!D^{\beta}_{0+}x(t) = \frac{p(t)}{M}.
\label{eq:FLE_x_frac_tau}
\end{align}
Under the same Fourier transform, with the Caputo derivative mapping to $(i\omega)^{\beta}$,
\begin{align}
\tau_{p}^{\,\beta-1}(i\omega)^{\beta}\,\widetilde{x}(\omega) = \frac{\widetilde{p}(\omega)}{M},
\label{eq:x_Fourier_alg}
\end{align}
and solving for $\widetilde{x}(\omega)$ expresses the position as,
\begin{align}
\widetilde{x}(\omega)
  = \frac{\widetilde{p}(\omega)}
         {M\,\tau_{p}^{\,\beta-1}(i\omega)^{\beta}}
  \equiv G(\omega)\,\widetilde{p}(\omega),
\qquad
G(\omega)=\frac{1}{M\,\tau_{p}^{\,\beta-1}(i\omega)^{\beta}}.
\label{eq:x_Fourier}
\end{align}
The position power spectral density follows from the same definition, $S_{x}(\omega)=\lim_{\mathcal{T}\to\infty}\tfrac{1}{\mathcal{T}}\langle|\widetilde{x}(\omega)|^{2}\rangle$. Substituting Eq.~\eqref{eq:x_Fourier} and taking the deterministic factor $|G(\omega)|^{2}$ outside from the average yields,
\begin{align}
S_{x}(\omega)
  = |G(\omega)|^{2}\,\lim_{\mathcal{T}\to\infty}\frac{1}{\mathcal{T}}
    \big\langle|\widetilde{p}(\omega)|^{2}\big\rangle
  = |G(\omega)|^{2}\,S_{p}(\omega),
\label{eq:Sx_Sp}
\end{align}
with $|G(\omega)|^{2}=1/[M^{2}\tau_{p}^{2(\beta-1)}|\omega|^{2\beta}]$ and $S_{p}(\omega)$ from Eq.~\eqref{eq:Sp_white},
\begin{align}
S_{x}(\omega)
  = \frac{2\mathcal{D}}
         {M^{2}\,\tau_{p}^{\,2(\beta-1)}\,|\omega|^{2\beta}\,
          \bigl|\gamma+\tau_{p}^{\,\alpha-1}(i\omega)^{\alpha}\bigr|^{2}}.
\label{eq:S_x}
\end{align}

For a process with stationary increments, the mean-square displacement is represented directly in terms of the generalized increment spectrum as~\cite{Yaglom1987,MandelbrotVanNess1968}
\begin{align}
\langle [x(t)-x(0)]^{2}\rangle
  = \frac{1}{2\pi}\int_{-\infty}^{\infty}\bigl|e^{i\omega t}-1\bigr|^{2}\,\mathcal{S}_{x}(\omega)\,d\omega,
\label{eq:msd_increment_spectral}
\end{align}
in which the factor $e^{i\omega t}-1$ arises because the observable is the displacement $x(t)-x(0)$ rather than the absolute position, and vanishes at $\omega=0$, removing the zero-frequency mode responsible for the non-stationarity. Using
\begin{align}
\bigl|e^{i\omega t}-1\bigr|^{2}
  = \bigl(e^{i\omega t}-1\bigr)\bigl(e^{-i\omega t}-1\bigr)
  = 2\bigl[1-\cos(\omega t)\bigr],
\label{eq:kernel_identity}
\end{align}
Eq.~\eqref{eq:msd_increment_spectral} becomes
\begin{align}
\langle [x(t)-x(0)]^{2}\rangle
  = \frac{1}{\pi}\int_{-\infty}^{\infty}\mathcal{S}_{x}(\omega)\,\bigl[1-\cos(\omega t)\bigr]\,d\omega,
\label{eq:msd_twosided}
\end{align}
and since $\mathcal{S}_{x}(-\omega)=\mathcal{S}_{x}(\omega)$ the integrand is even, so that
\begin{align}
\langle [x(t)-x(0)]^{2}\rangle
  = \frac{2}{\pi}\int_{0}^{\infty}\mathcal{S}_{x}(\omega)\,\bigl[1-\cos(\omega t)\bigr]\,d\omega.
\label{eq:msd_general}
\end{align}
The particle is taken to start at the origin, $x(0)=0$, so that $\langle[x(t)-x(0)]^{2}\rangle=\langle x^{2}(t)\rangle$,
\begin{align}
\langle x^{2}(t)\rangle
  = \frac{2}{\pi}\int_{0}^{\infty}
    S_{x}(\omega)\,\bigl(1-\cos\omega t\bigr)\,d\omega.
\label{eq:msd_general}
\end{align}
Substituting Eq.~\eqref{eq:S_x} with the expanded modulus of Eq.~\eqref{eq:denom_expand} yields the integral representation
\begin{align}
\langle x^{2}(t)\rangle
  = \frac{4\mathcal{D}}
         {\pi M^{2}\,\tau_{p}^{\,2(\beta-1)}}
    \int_{0}^{\infty}
    \frac{(1-\cos\omega t)\,d\omega}
         {\omega^{2\beta}\,
          \bigl[\gamma^{2}
                +2\gamma\,\tau_{p}^{\,\alpha-1}\omega^{\alpha}\cos\tfrac{\pi\alpha}{2}
                +\tau_{p}^{\,2(\alpha-1)}\omega^{2\alpha}\bigr]}.
\label{eq:msd_exact}
\end{align}
At late times $t\gg\tau_{p}$ the integral is dominated by the small-$\omega$ region, where the bracket reduces to $\gamma^{2}$ and the integrand simplifies to $(1-\cos\omega t)/\omega^{2\beta}$. Using the standard identity, valid for $1<\beta<\tfrac{3}{2}$,
\begin{align}
\int_{0}^{\infty}
\frac{1-\cos\omega t}{\omega^{2\beta}}\,d\omega
=
\frac{\pi}{2\,\Gamma(2\beta)\,\sin\!\big[\pi\big(\beta-\tfrac12\big)\big]}\,t^{\,2\beta-1},
\label{eq:cosine_identity}
\end{align}
in which $\sin[\pi(\beta-\tfrac12)]>0$ on this interval, the long-time mean-square displacement is
\begin{align}
\langle x^{2}(t)\rangle
  \;\sim\;
  \frac{2\mathcal{D}}
       {M^{2}\,\gamma^{2}\,\tau_{p}^{\,2(\beta-1)}}\,
  \frac{t^{\,2\beta-1}}
       {\Gamma(2\beta)\,\sin\!\big[\pi\big(\beta-\tfrac12\big)\big]},
\qquad
1<\beta<\tfrac{3}{2},\;\; t\gg\tau_{p}.
\label{eq:msd_asymptotic}
\end{align}
Equation~\eqref{eq:msd_asymptotic} exhibits the anomalous power law $\langle x^{2}(t)\rangle\propto t^{\,2\beta-1}$. In the limit $\alpha\to1$ and $\beta\to1$, $K_{\alpha}\to1$, the combined memory-time exponent gives $\tau_{p}^{(\alpha-1)/\alpha-2(\beta-1)}\to1$, and $\Gamma(2\beta)\,\sin[\pi(\beta-\tfrac12)]\to\Gamma(2)\sin(\tfrac{\pi}{2})=1$, reducing Eq.~\eqref{eq:msd_asymptotic} to
\begin{align}
\langle x^{2}(t)\rangle
  \;\to\;
  2\,D_{s}\,t,
\qquad
D_{s}\equiv\frac{T}{M\gamma},
\label{eq:msd_markovian}
\end{align}
in agreement with the fluctuation dissipation theorem of the standard Langevin dynamics of heavy-quark in the QGP.


\section{Numerical scheme for the fractional Langevin equation}

The numerical difficulty resides entirely in the Caputo operator. Because it is nonlocal in time, the increment at step $n$ depends on the whole preceding trajectory rather than on the current state alone, so the standard first-order integrators do not carry over to fractional order. We therefore discretize the operator directly via the L2 scheme~\cite{MR361633}, in which a centred
second difference is weighted by the fractional kernel and the history is retained through the coefficients $b_j$.


We solve the FLE for the HQ momentum and position numerically. As a consistency check, numerical treatments of time-fractional differential equations are generally grouped into two categories. Indirect schemes convert the problem into an equivalent integro-differential form and then solve that equation. Direct schemes, by contrast, approximate the fractional derivative itself without performing such a reformulation. We adopt the direct approach here: our algorithm discretizes the fractional derivative using a three-level, central-difference L2 scheme for the momentum and position of the HQs, as outlined in Appendix~\ref{AP}, and it is written as follows,

\begin{align}\label{SuperA_x}
\begin{cases}
x(t_1)= x(t_0) + k_{0}\displaystyle\frac{p(t_0)}{E(t_0)} &:\; n=1, \\[4pt]
x(t_2)= 2x(t_1)-x(t_0) -\left(2^{2-\beta}-1\right)\left[x(t_1)-x(t_0)\right] + k_{0}\displaystyle\frac{p(t_1)}{E(t_1)} &:\; n=2, \\[4pt]
x(t_n)= 2x(t_{n-1})-x(t_{n-2}) + k_{0}\displaystyle\frac{p(t_{n-1})}{E(t_{n-1})} & \\
\qquad -\,k\displaystyle\sum_{j=1}^{n-2}b_j\left[x(t_{n-j})-2x(t_{n-j-1})+x(t_{n-j-2})\right] & \\
\qquad -\left[n^{2-\beta}-(n-1)^{2-\beta}\right]\left[x(t_1)-x(t_0)\right] &:\; n \geq 3,
\end{cases}
\end{align}
and
\begin{align}\label{SuperA_p}
\begin{cases}
p(t_1)= p(t_0) + k_0\left[-\gamma p(t_0)+\sqrt{2\mathcal{D}/\Delta t}\,\eta\right] &:\; n=1, \\[4pt]
p(t_2)= 2p(t_1)-p(t_0) -\left(2^{2-\alpha}-1\right)\left[p(t_1)-p(t_0)\right] + k_0\left[-\gamma p(t_1)+\sqrt{2\mathcal{D}/\Delta t}\,\eta\right] &:\; n=2, \\[4pt]
p(t_n)= 2p(t_{n-1})-p(t_{n-2}) + k_0\left[-\gamma p(t_{n-1})+\sqrt{2\mathcal{D}/\Delta t}\,\eta\right] & \\
\qquad -\,k\displaystyle\sum_{j=1}^{n-2}b_j\left[p(t_{n-j})-2p(t_{n-j-1})+p(t_{n-j-2})\right] & \\
\qquad -\left[n^{2-\alpha}-(n-1)^{2-\alpha}\right]\left[p(t_1)-p(t_0)\right] &:\; n \geq 3,
\end{cases}
\end{align}
where

$k_0 = \Delta t^{\alpha}\Gamma(3-\alpha)/\tau_{p}^{\,\alpha-1}$, and $k = \Delta t^{\alpha}\Gamma(3-\alpha)$  with $t_{1}= t_{0} + \Delta{t}$, $t_{n}= t_{n-1} +\Delta t$. Here $t_{0}$ is the initial time, and $\Delta{t}$ is the time step. In the calculations presented in this work, $\alpha=\beta$.

\begin{figure}[h]
\centering
\includegraphics[scale=0.43]{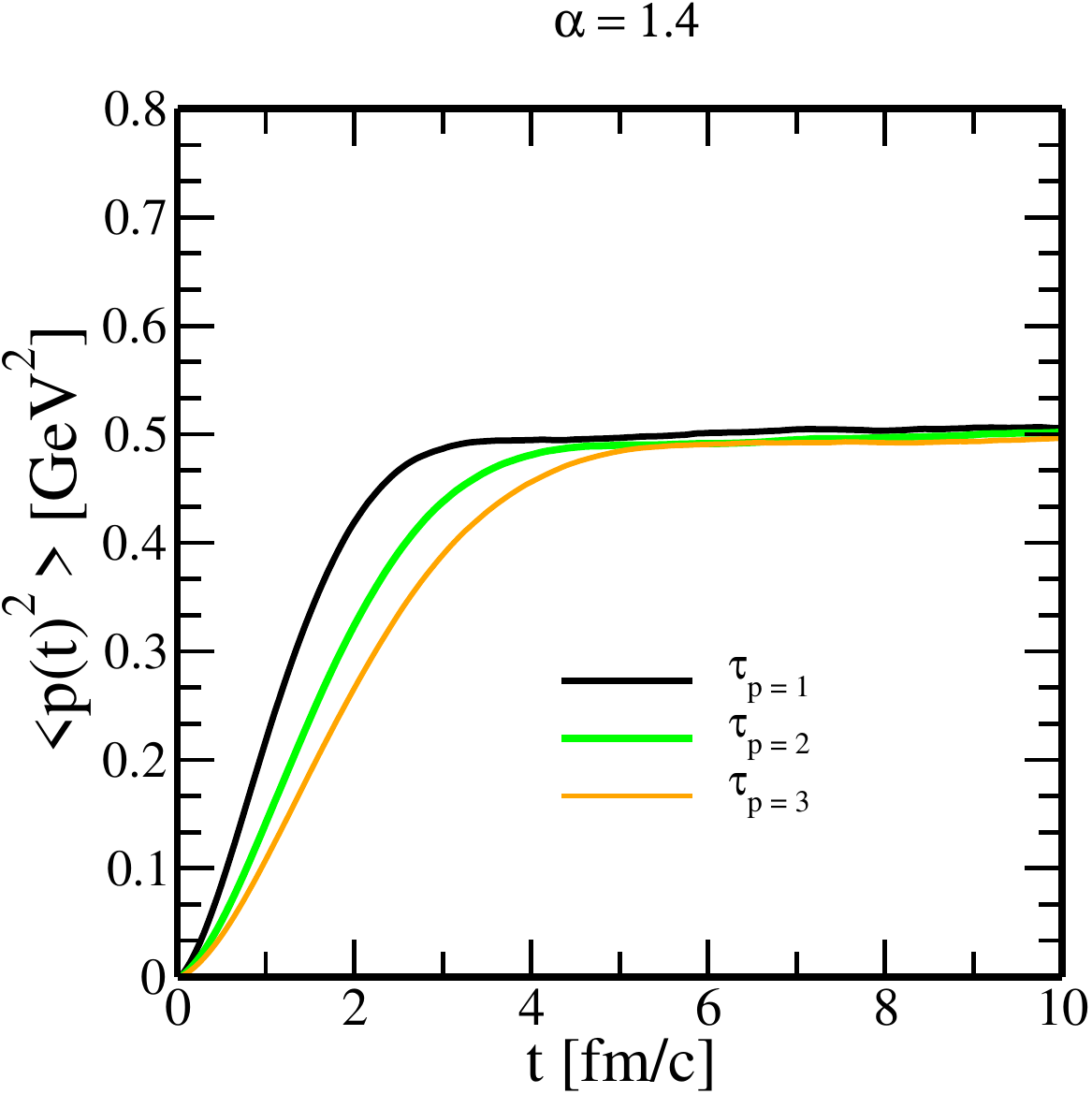}
\caption{Time evolution of $\langle p^{2}(t)\rangle$ of a Brownian particle in the non-relativistic limit for three values of $\tau_{p}=1,\,2,\,3~\mathrm{fm}/c$, at fixed fractional order $\alpha=1.4$, $M=1.0~\mathrm{GeV}$ and $T = 0.25$~GeV.}
\label{fig:pt2_taup}
\end{figure}

Before turning to the systematic study of the $\alpha$-dependence, it is reasonable to show the effect of the  $\tau_{p}$  on the HQ dynamics in the QGP medium. We show in Fig.~\ref{fig:pt2_taup} the time evolution of the  $\langle p^{2}(t)\rangle$ of a Brownian particle, computed at a fixed fractional order $\alpha=1.4$, for three representative values $\tau_{p}=1$, $2$, and $3~\mathrm{fm}/c$.  The three curves display the same asymptotic behaviour: at late times they all approach a common plateau, set by the equilibrium relation $\langle p^{2}\rangle_{\rm eq}=2MT$ in the two-dimensional non-relativistic limit, with no dependence on $\tau_{p}$. The differences are confined to the early-time regime: smaller values of $\tau_{p}$ lead to a faster initial rise of $\langle p^{2}(t)\rangle$, whereas larger values delay the buildup. The independence of the late-time plateau on $\tau_{p}$ shows that the HQ thermal condition is independent of the value of $\tau_{p}$. In this work, $\tau_{p}$ is introduced solely on dimensional grounds, to render the modified F-FDT dimensionally consistent for $\alpha\neq1$, and not as a physical parameter controlling thermalisation.

\section{Caputo derivative: L2 discretization}

\label{AP}

Divide the interval $[0,T]$ into $N$ equal subintervals by introducing the grid points
$t_n = \frac{nT}{N}$ for $0 \le n \le N$. In the superdiffusive case, one can employ the
three-step L2 fractional scheme. This method uses a centred finite-difference approximation
for the second derivative, and the resulting discrete formula at step $n$ involves
the previous steps.

For $1<\nu\leq 2$, the Caputo derivative at $t_n$ is split over the grid intervals
$[t_{k-1},t_k]$, $1\leq k\leq n$, with $\Delta t=t_k-t_{k-1}$. The kernel
integral is evaluated on each interval and the second derivative is represented by
the three-point finite difference:
\begin{align}
\nonumber {}^{C}D^\nu_{0+}u(t_n)
&=\frac{1}{\Gamma(2-\nu)}
\int_0^{t_n}\frac{u^{(2)}(s)}{(t_n-s)^{\nu-1}}\,ds\\
\nonumber
&=\frac{1}{\Gamma(2-\nu)}
\sum_{k=1}^{n}\int_{t_{k-1}}^{t_k}
\frac{u^{(2)}(s)}{(t_n-s)^{\nu-1}}\,ds\\
\nonumber
&\approx\frac{1}{\Gamma(2-\nu)}
\sum_{k=1}^{n}
\frac{u(t_k)-2u(t_{k-1})+u(t_{k-2})}{\Delta t^2}
\int_{t_{k-1}}^{t_k}(t_n-s)^{1-\nu}\,ds\\
\nonumber
&=\frac{1}{\Gamma(2-\nu)}
\sum_{k=1}^{n}
\frac{u(t_k)-2u(t_{k-1})+u(t_{k-2})}{\Delta t^2}
\left[-\frac{(t_n-s)^{2-\nu}}{2-\nu}\right]_{t_{k-1}}^{t_k}\\
\nonumber
&=\frac{1}{\Gamma(3-\nu)}
\sum_{k=1}^{n}
\left[(t_n-t_{k-1})^{2-\nu}-(t_n-t_k)^{2-\nu}\right]
\frac{u(t_k)-2u(t_{k-1})+u(t_{k-2})}{\Delta t^2}.
\end{align}
On a uniform grid, the coefficient of the $k$-th interval depends on $n$ and
$k$ only through the time lag $n-k$. We therefore introduce $j=n-k$. Thus, $j=0$ denotes
the most recent interval and increasing $j$ labels progressively earlier
intervals. The discrete derivative then becomes
\begin{align}
{}^{C}D^\nu_{0+}u(t_n)\approx{}&
\frac{1}{\Gamma(3-\nu)}
\sum_{j=0}^{n-1}
\left[(t_n-t_{n-j-1})^{2-\nu}-(t_n-t_{n-j})^{2-\nu}\right]
\nonumber\\
&\times
\frac{u(t_{n-j})-2u(t_{n-j-1})+u(t_{n-j-2})}{\Delta t^2}.
\label{eq:L2_shifted}
\end{align}
On a uniform grid, $t_n-t_{n-j}=j\Delta t$, and the weight associated with
the $j$-th interval is
$\Delta t^{2-\nu}\bigl[(j+1)^{2-\nu}-j^{2-\nu}\bigr]$. We therefore define
\begin{align*}
b_j&=\frac{(j+1)^{2-\nu}-j^{2-\nu}}
{\Gamma(3-\nu)\Delta t^\nu},
&
b_0&=\frac{1}{\Gamma(3-\nu)\Delta t^\nu}.
\end{align*}
For $1<\nu\leq2$, two initial values are required. We define
$u(t_0)=u_0$, with $t_0=0$, and take the initial linear contribution
to vanish. For the stochastic problem, this is imposed as an initial condition and does not require the existence of a classical pathwise derivative $u'(t_0)$. For $j=n-1$, the second difference in Eq.~\eqref{eq:L2_shifted}
contains $u(t_{-1})$, where $t_{-1}=-\Delta t$ lies outside the
computational domain. The zero initial linear condition is implemented
discretely as
\[
\frac{u(t_0)-u(t_{-1})}{\Delta t}=0,
\]
which yields $u(t_{-1})=u(t_0)$. Hence, on the initial interval,
\[
u(t_1)-2u(t_0)+u(t_{-1})
=
u(t_1)-u(t_0).
\]
The complete three-point formula is used from $n=3$ onward; the
first two values are obtained from
\begin{align*}
{}^{C}D^\nu_{0+}u(t_1)
&\approx\frac{u(t_1)-u(t_0)}{\Gamma(3-\nu)\Delta t^\nu},\\
{}^{C}D^\nu_{0+}u(t_2)
&\approx\frac{u(t_2)-2u(t_1)+u(t_0)+\left(2^{2-\nu}-1\right)\left[u(t_1)-u(t_0)\right]}{\Gamma(3-\nu)\Delta t^\nu}.
\end{align*}
The complete discrete formula is
\begin{align}\label{SuperA}
{}^{C}D^\nu_{0+}u(t_n)\approx
\begin{cases}
\displaystyle
\frac{u(t_1)-u(t_0)}{\Gamma(3-\nu)\Delta t^\nu},
& n=1,\\[8pt]
\displaystyle
\frac{u(t_2)-2u(t_1)+u(t_0)+\left(2^{2-\nu}-1\right)\left[u(t_1)-u(t_0)\right]}{\Gamma(3-\nu)\Delta t^\nu},
& n=2,\\[8pt]
\begin{aligned}
&\displaystyle
\frac{u(t_n)-2u(t_{n-1})+u(t_{n-2})}
{\Gamma(3-\nu)\Delta t^\nu}\\
&\displaystyle\quad+
\sum_{j=1}^{n-2}b_j
\left(u(t_{n-j})-2u(t_{n-j-1})+u(t_{n-j-2})\right)\\
&\displaystyle\quad+
\frac{n^{2-\nu}-(n-1)^{2-\nu}}
{\Gamma(3-\nu)\Delta t^\nu}\left[u(t_1)-u(t_0)\right],
\end{aligned}
& n\geq3.
\end{cases}
\end{align}
The coefficients $b_j$ weight the contributions from the earlier time intervals.

\printbibliography